\documentclass[preprint,aps,11pt,showpacs,showkeys,nofootinbib]{revtex4-1}

\usepackage{amssymb}
\usepackage{graphicx}
\usepackage{epstopdf}    
\usepackage{dcolumn}
\usepackage{amsfonts}
\usepackage{amsmath}
\usepackage{amssymb}
\usepackage{bm}
\usepackage{color}
\usepackage{comment}
\newcolumntype{L}{>{\centering\arraybackslash}m{1.9cm}}

\newcommand{\be}{\begin{equation}}
\newcommand{\lab}{\label}
\newcommand{\ee}{\end{equation}}
\newcommand{\bea}{\begin{eqnarray}}
\newcommand{\eea}{\end{eqnarray}}
\newcommand{\bqa}{\begin{eqnarray}}
\newcommand{\eqa}{\end{eqnarray}}
\newcommand{\bwt}{\begin{widetext}}
	\newcommand{\ewt}{\end{widetext}}

\newcommand{\nn}{\nonumber \\}

\newcommand{\del}{\delta}

\newcommand{\ti}{\tilde}

\newcommand{\bfs}{{\bf S}}
\newcommand{\re}[1]{(\ref{#1})}
\newcommand{\ba}{\begin{array}{l}   }
\newcommand{\ea}{\end{array}{l}   }

\begin{document}


\title{Spin-gapped magnets with weak anisotropies \\
	  I: Constraints on the phase of the condensate wave function}

\author{Abdulla Rakhimov$^{a}$}\email{rakhimovabd@yandex.ru}
\author{Asliddin Khudoyberdiev $^{b}$}\email{asliddinkh@gmail.com}
\author{Luxmi Rani$^{a}$}\email{luxmi.rani@bilkent.edu.tr}
\author{B. Tanatar $^{a}$}\email{tanatar@fen.bilkent.edu.tr}	

\affiliation{$^a$Department of Physics, Bilkent University, Bilkent, 06800 Ankara, Turkey}
\affiliation{$^b$Institute of Nuclear Physics, Tashkent 100214, Uzbekistan}
\date{\today}
\begin{abstract}
We study the thermodynamic properties of dimerized spin-gapped quantum magnets with and without exchange anisotropy (EA) and Dzyaloshinsky and Moriya (DM) anisotropies within the mean-field approximation (MFA). For this purpose we obtain the thermodynamic potential $\Omega$ of a triplon gas taking into account the strength of DM interaction up to second order. The minimization of $\Omega$ with respect to self-energies $\Sigma_n$ and $\Sigma_{an}$ yields the equation for $X_{1,2}=\Sigma_n\pm\Sigma_{an}-\mu$, which define  the dispersion of quasiparticles $E_k=\sqrt{\epsilon_k+X_1}\sqrt{\epsilon_k+X_2}$ where $\epsilon_k$ is the bare dispersion of triplons. The minimization of $\Omega$ with respect to the magnitude $\rho_0$ and the phase $\Theta$ of triplon condensate leads to coupled equations for $\rho_0$ and
$\Theta$. We discuss the restrictions on $\rho_0$ and $\Theta$ imposed by these equations  for systems with and without anisotropy. The requirement of dynamical stability conditions $(X_1>0, X_2>0)$ in equilibrium, as well as the Hugenholtz-Pines theorem, particularly for isotropic Bose condensate, impose certain conditions to the physical solutions of these equations. It is shown that the phase angle of a purely homogenous Bose-Einstein condensate (BEC) without any anisotropy may only take values $\Theta=\pi n $ (n=0,$\pm 1,\pm 2$...) while that of BEC with even a tiny DM interaction results in $\Theta=\pi/2+2\pi n$. In contrast to the widely used Hartree-Fock-Popov approximation, which allows arbitrary phase angle, our approach predicts that the phase angle may have only discrete values,
while the phase of the wave function of the whole system remains arbitrary as expected.
The consequences of this phase locking for interference of two Bose condensates and to their possible
Josephson junction is studied.
In such quantum magnets the emergence of a triplon condensate leads to a finite staggered magnetization $M_\bot$, whose direction in the xy-plane is related to the condensate phase $\Theta$. We also discuss the possible Kibble-Zurek mechanism in dimerized magnets and its influence on $M_\bot$.

\end{abstract}





	\maketitle

\section{Introduction}
Triplons are bosonic quasi-particles introduced in bond operator formalism \cite{Sachdev} to describe the singlet-triplet excitations in spin-gapped magnetic materials. Measurement of the magnetization of such antiferromagnetic compounds have shown \cite{2} an interesting dependence of magnetization $M(T)$ at low temperatures: $M(T)$ decreases with decreasing temperature and unexpectedly starts to increase when temperature becomes lower than a critical temperature, $T_c$. Moreover, it is observed that such behavior of $M(T)$ can take place only when the external magnetic field exceeds a critical value $H_c$, i.e., $(H \geq H_c)$. Further measurements based on inelastic neutron scattering \cite{Ruzfber} have revealed that in some compounds such as KCuCl$_3$ or TlCuCl$_3$ two Cu$^{2++}$ ions are antiferromagnetically coupled to form a dimer in crystalline network. There is a gap $\Delta_{st} \approx 0.6$\,meV between the ground singlet (S=0) and excited triplet (S=1) states, which can be closed owing to the Zeeman effect for
$H \geq H_c$ with $\Delta_{st}= g\mu_B H_c$, where Land{\'e} electron $g$-factor is $g\approx 2$ and
Bohr magneton is $\mu_B = 0.672$\,K/T. Subsequently, the list of spin-gapped magnets has been extended as reviewed in Ref.\,\cite{Zapf}.

In the ideal case, the triplons have axial symmetry O(3) which can be spontaneously broken leading to a Bose-Einstein condensation (BEC), similar to a BEC of atoms arising from the spontaneous breaking of U(1) symmetry. Now, relating the uniform magnetization $M$ to the total number of triplons as $M= g\mu_B N$, and the staggered magnetization to the condensate fraction as $M_\perp =g\mu_B \sqrt{N_0/2}$, one may describe experimental data on magnetization \cite{Zapf}. As to the data on energy dispersion of collective excitations, R{\"u}egg \emph{et al.} \cite{Ruegg} showed that at low temperatures the spectrum becomes gapless and they may be naturally explained by the existence of a Goldstone mode in a BEC.

Further neutron scattering experiments \cite{Tanasca2001} have shown that the staggered magnetization of dimerized spin gap system of  TlCuCl$_3$ remains finite even at $T> T_c$. This contradicts a pure BEC model based on spontaneous symmetry breaking, when, by definition, $N_0 (T >T_c)=0$. Therefore, to improve the situation one may suppose that the rotational (axial) symmetry is weakly explicitly broken in this sample.  In fact, in real systems, there may always be weak anisotropy, breaking O(3) symmetry, such as crystalline anisotropies, spin-orbit coupling and dipole interactions. Even when they are weak, the anisotropies may become important at low temperatures and modify the physical properties \cite{Iamarchi}.

In general, the exchange interaction between two moments has the form $\Sigma_{ij}S_r^iT^{ij}S_{r+e_\nu}^j$ \cite{Sebastian2006} where $\textbf{T}=(1/3)Tr(\textbf{T})I+\textbf{T}_{as}+\textbf{T}_{SM}$. Here the first term leads to the usual isotropic exchange coupling, $\textbf{T}_{as}$ is an antisymmetric tensor that describes the Dzyaloshinsky-Moriya (DM) interaction $\textbf{D}\cdot [\textbf{S}_r\textbf{S}_{r+e_\nu}]$, where $\textbf{D}$ is the DM vector. The last term contains the so-called symmetric exchange anisotropy (EA) and has contributions from the classical dipole-dipole interaction between magnetic moments.
Clearly, such interactions break axial symmetry not spontaneously but explicitly.

Spontaneous symmetry breaking (SSB) was originally developed to explain spontaneous magnetization in ferromagnetic systems. Spontaneous coherence in all its forms can be viewed as another type of symmetry breaking. In SSB the Hamiltonian of the system is symmetric, yet under some conditions, the energy of the system can be reduced by putting the system into a state with asymmetry, namely a state with a common phase for a macroscopic number of particles. The symmetry of the system implies that it does not matter what the exact choice of that phase is, as long as it is the same for all particles \cite{snoke}.
Therefore, when the system of triplons is invariant under rotational symmetry, one deals with spontaneous symmetry breaking and hence with pure BEC is termed as isotropic. On the other hand, when DM or EA interactions exist, one has to deal with an explicitly broken axial symmetry where strictly speaking no BEC can take place\cite{yukalovehaya}. Nevertheless, due to the weakness of observed anisotropy we shall apply BEC model calling it the anisotropic case, and study its consequences for physical observables.

A similar study has been performed for the first time by Sirker \emph{et al}. \cite{Sirker2}. They attempted to describe the experimental data \cite{2,Tanasca2001} on TlCuCl$_3$ and came to an important conclusion: the inclusion of DM interaction into the standard Hamiltonian with a contact interaction smears out the phase transition into a crossover. In other words, in the presence of DM anisotropy a critical temperature $T_c$ (defined as $(dN/dT)_{T=T_c}=0$) may still exist, but the condensate fraction $N_0(T)$ subsides asymptotically with increasing temperature. The effect of the presence of only EA interaction has been also studied within BEC concept \cite{23 our aniz, Delamore}. Delamore \emph{et al}. \cite{Delamore} based on an energetic argument on classical level predicted a general intrinsic instability of triplon condensate due to EA.
It is interesting to know if this prediction remains true when quantum fluctuations are taken into account. As to the works by Sirker \emph{et al}. {\cite{Sirker2, Sirker1}} there are two main points we wish to improve.
(i) The anomalous density, $\sigma$ has been neglected. This corresponds to the mean field approach (MFA) called in the literature as Hartree-Fock-Popov (HFP) approximation
\footnote{{Strictly speaking, it would not be correct to associate  this approximation with the name of V. N. Popov, since as we know, he never used or suggested it.}}
 and predicts  an unexpected cusp in magnetization near $T_c$ in the pure BEC case for compounds whose anisotropies are negligibly small \cite{yamada}. Further, it was shown that \cite{our papers,ourANN} application of Hartree-Fock-Bogoliubov  (HFB) approximation, which includes anomalous density may improve the theoretical description.
(ii) The DM interaction with intensity $\gamma'$, has been taken into account up to the first order in $\gamma'$ as a perturbation. However in reality, it is well known that among the magnetic anisotropies,  DM interaction is the strongest and dominate over e.g., the EA interaction.

The main goal of the present work is to study the low temperature properties of spin gapped magnets with DM and exchange anisotropies by taking into account their contribution more systematically than in Refs.\, {\cite{23 our aniz,Delamore,Sirker1}}  including anomalous density. For this purpose, we shall represent DM and EA interactions by a linear (in fields) and quadratic terms, respectively, and use $\delta$-expansion method developed in quantum field theory \cite{stancu, zamos, chiku}. We will show that the present approach gives a better description for the proper choice of the phase of the condensate.

 The phase of the condensate plays an important role in interference experiments, where two BECs are released and resulting interference patterns are measured \cite{andrews}. However, in the experiments performed, it is unknown whether the condensate has a well-defined phase before the measurement was made or only afterward. We show that a BEC of an noninteracting system has an arbitrary, random phase, while the presence of interparticle interactions make BEC to acquire a preferred phase which guaranties its stability. Then, we study the consequences of this conclusion for some physical phenomena such as Josephson junctions and Kibble-Zurek mechanism (KZM).
In the present work we suggest a method for observing KZM in dimerized magnets. Namely, we propose experimentally to make a rapid quench in a spin gapped quantum magnet below $T<T_c$ and measure the staggered magnetization
$M_\perp$. We predict that for a material with even a tiny DM anisotropy, $M_\perp$ is not sensitive to the time of quenching, while that of a pure magnet without DM anisotropy vanishes in ``rapid quenching'' and remains finite in
``slow quenching''.

The rest of this article is organized as follows. In section II, we present the thermodynamic potential $\Omega$, whose detailed derivation is moved to Appendix \ref{sec:A}. In section III, we discuss the phase of the condensate and properties of our equations with respect to self-energies; in section \textrm{IV}, we discuss interference and Josephson junction of two Bose systems and then we study the KZM in spin gapped magnets in section \textrm{V}. The section \textrm{VI} summarizes our main conclusions.

\section{The thermodynamic potential including EA and DM interactions}

In bond operator formalism for magnetic fields greater than the critical field, $ H\geq H_c $, the Hamiltonian of a triplon gas with exchange and DM anisotropies can be presented as the sum of ``isotropic" and ``anisotropic" terms
\footnote{
See Appendix A for some details.
}
\begin{subequations}
	\begin{align}
	{\cal H}=&H_{iso}+H_{aniso}, \label{eq:H}\\
	H_{iso}=&\int d\vec{r}\left[\psi^+(r) (\hat{K}-\mu) \psi(r)+ \frac{U}{2} (\psi^+(r)\psi(r))^2\right], \label{eq:H1}\\
	H_{aniso}=&H_{EA}+ H_{DM},\label{eq:H2}\\
	H_{EA}=&\frac{\gamma}{2} \int d\vec{r} \left[\psi^+(r)\psi^+(r) + \psi(r)\psi(r) \right], \label{eq:H3} \\
	H_{DM}= &i\gamma'\int d\vec{r} \left[\psi(r)-\psi^+(r) \right],
	\label{eq:H4}
	\end{align}
\end{subequations}
where $\psi(r)$ is the bosonic field operator
of a quasiparticle - triplon,
 $U, \gamma, \gamma'$ are the interaction strengths
($U\geq 0, \gamma \geq 0, \gamma' \geq 0$) and $\hat{K}$ is the kinetic energy operator which defines the bare triplon dispersion $\varepsilon_k$ in momentum space.\footnote{Here and below we adopt the units $k_B=1$ for the Boltzmann constant, $\hbar=1$ for the Planck constant, and V=1 for the unit cell volume.} The integration is performed over the unit cell of the crystal with corresponding momenta defined in the first Brillouin zone \cite{23 our aniz}. The parameter $\mu$ characterizes an additional direct contribution to the triplon energy due to the external magnetic field $H$,
\bea
\mu= g \mu_B (H-H_c) \equiv g \mu_B H -\Delta_{st}
\eea
and it may be interpreted as the chemical potential of the $ S_z = -1 $ triplons. The spin gap $\Delta_{st}$  separating the singlet ground state from the lowest-energy triplet excitations defines the critical field $H_c$, above which it is closed by Zeeman splitting. The linear Hamiltonian, an external source, in Eq.\,\eqref{eq:H4} corresponds to a simple case when singlet-triplet mixing is neglected and DM vector is chosen as $D \parallel x$ and $H\parallel z$ \cite{Sirker1}. It is often used as an artificial interaction with $\gamma'\rightarrow0$ in quantum field theories to derive exact relations such as Ward-Takahashi identities \cite{Enomoto}. However, in the present work we assume $\gamma'$ to be small but finite to study its physical consequences. A more general and complicated expression for $H_{DM}$ can be found e.g., in Ref. \cite{Miyahara}. Note that, $H_{iso}$ is symmetric under gauge transformation $\psi \rightarrow e^{i\phi} \psi$, while $H_{aniso}$ is not. Therefore, strictly speaking, there would be neither a Goldstone mode nor a pure Bose condensation \cite{yukalovehaya, 23 our aniz}. Nevertheless, assuming $ \gamma/U \ll 1$ and $ \gamma'/U \ll 1$, one may separate the condensate contribution, which corresponds to the macroscopic occupation of a single quantum state, from the remaining part of the Bose field operator. Therefore, assuming that just one state of the system to be occupied macroscopically, it is natural to re-arrange the Bose field operator into two parts
\bea
\psi(r, t)= \chi (r, t) + \ti{\psi}(r, t), \quad  \psi^+(r, t)= \chi^+ (r, t) + \ti{\psi}^+(r, t),
\label{eq:psi}
\eea
corresponding, respectively, to a field operator for the condensate  $\chi(r,t)$ and one for the non-condensed particles
$\ti{\psi}$. These could correspond, mainly, to thermal excitations and to quantum fluctuations. As to the operator $\chi$, for a homogeneous system it is usually called the condensate wave function which is a complex number including the order parameter $\rho_0 $ and the phase $\xi$ of the condensate:
\bea
\chi \equiv \xi \sqrt{\rho_0} \equiv e^{i\Theta} \sqrt{\rho_0}, \quad
\chi^+ \equiv \xi^+\sqrt{\rho_0} \equiv e^{-i\Theta} \sqrt{\rho_0}.
\label{eq:chi}
\eea
In an equilibrium system, the condensate wave function does not depend on time, $\chi(r,t)\equiv\chi(r)$.
The split in Eq.\,(\ref{eq:psi}) clearly demonstrates the SSB. In fact, even when the system Hamiltonian is invariant under a gauge transformation in the phase of $\psi(r, t)$, ($\gamma=\gamma'=0$ case), the wave function $\chi$ no longer shares this symmetry. In other words, after insertion of Eq.\,(\ref{eq:psi}) into the Hamiltonian $H_{iso}$, it will be invariant with respect to $\psi \rightarrow e^{i\phi}\psi$ transformation only in the normal phase ($T>T_c$) where $\chi =0$.\footnote{Further, for simplicity, we shall use notation $\bar{\chi}\equiv \chi^+$, $\bar{\xi}\equiv \xi^+$}

In equilibrium the free energy $\Omega$ reaches its minimum :
\begin{subequations}
	\begin{align}
	\frac{\partial \Omega}{\partial \rho_0}&=0, \quad  \frac{\partial^2 \Omega}{\partial \rho_{0}^{2}}\geq 0.\label{eq:partial1}\\
	\frac{\partial \Omega}{\partial \Theta}&=0, \quad  \frac{\partial^2 \Omega}{\partial \Theta^{2}}\geq 0.
	\label{eq:partial2}
	\end{align}
\end{subequations}

As it was shown by Andersen \cite{andersen} Eqs.\,(5) are equivalent to satisfying the quantum number conservation condition such as $\langle\ti{\psi}(r)\rangle=0$, $\langle\ti{\psi}^+(r)\rangle=0$. \cite{Yukalov review}.
Here it should be underlined that, Eqs. \re{eq:partial2} impose a constraint to the phase of only 
the condensate wave function $\chi$, while the phase of the  $\ti{\psi}$ in \re{eq:psi}, and therefore 
that of the  wave function, corresponding to 
the operator $\psi$ remains arbitrary in accordance with general laws of quantum mechanics.

One of our main goals is to find an analytical expression for $\Omega$, which contains almost all the information about the equilibrium statistical system. For example, entropy and magnetization may be evaluated as $ S=-(\partial \Omega/ \partial T)$, $M=-(\partial \Omega/ \partial H)=g\mu_B N$ from the thermodynamic relation \cite{ourjt}:
\bea
d\Omega = - SdT - PdV - Nd\mu - MdH.
\label{eq:domega}
\eea
For this purpose we use the path integral formalism where $\Omega$ is given by
\begin{subequations}\label{eq:entropy}
	\begin{equation}	
	\Omega = - T\ln{Z},
	\tag{\ref{eq:entropy}}
	\end{equation}
	\begin{align}
	Z & = \int  D \ti\psi^+ D \ti\psi e^{-A[\psi,\psi^+]}, \\
	A[\psi,\psi^+]&= \nn
	& \int_{0}^{\beta}d{\tau} d\vec{r} \left\lbrace \psi^+ \left[\frac{\partial}{\partial\tau}-\hat{K}-\mu\right] \psi + \frac{U}{2}\left( \psi^+\psi\right)^2  +\frac{\gamma}{2}\left( \psi\psi+\psi^+\psi^+\right)
 + i \gamma'\left( \psi-\psi^+\right)\right\rbrace.
   \label{eq:entropy1}
	\end{align}
\end{subequations}
In Eq.\,\eqref{eq:entropy1} the fluctuating fields $\ti{\psi}(r, \tau)$ and $\ti{\psi}^+(r, \tau)$ satisfy the bosonic commutation relations and are periodic in $\tau$ with period $\beta=1/T$. Clearly, this path integral can not be evaluated exactly, so an approximation is needed. In the present work, we shall use an approach, which is called the variational perturbation theory \cite{stancu,zamos, ourKL,KLbook}, or $\delta$-expansion method. We apply this method as follows (see Appendix A for details):\\
(1) Make following replacements in the action \eqref{eq:entropy1}: $U\rightarrow \delta U$, $\gamma\rightarrow \delta \gamma$, $\gamma'\rightarrow \sqrt{\del}\gamma' $.\\
(2) Add to the action the term:
\bea
A_\Sigma= (1-\delta)\int d{\tau} d\vec{r}\left[ \Sigma_n\ti{\psi^+}\ti{\psi} + \frac{1}{2} \Sigma_{an}\left(\ti{\psi^+}\ti{\psi^+}+\ti{\psi}\ti{\psi} \right) \right]
\label{eq:Asigma}
\eea
where the variational parameters $\Sigma_n$ and $\Sigma_{an}$ may be interpreted as the normal and anomalous
self-energies, respectively. They are defined as \cite{andersen}:
\begin{subequations}
	\begin{align}
	\Sigma_{n}=(\Pi_{11}(0,0)+\Pi_{22}(0,0))/2,\\ \Sigma_{an}=(\Pi_{11}(0,0)-\Pi_{22}(0,0))/2, \\ \Pi_{ab}(\omega_{n},\vec{k})=(G(\omega_{n},\vec{k}))^{-1}_{ab}-(G^{0}(\omega_{n},\vec{k}))^{-1}_{ab}
	\label{eq:9}
	\end{align}
\end{subequations}
with the Green functions $G(\omega_n, \vec{k})$, $G^{0}(\omega_n, \vec{k})$  given below.\\
(3) Now the perturbation scheme may be considered as an expansion in powers of $\delta$ by using the propagators
\bea
G_{ab}(\tau, \vec{r}; \tau', \vec{r}')=\frac{1}{\beta}\sum_{n,k} e^{i\omega_n(\tau-\tau')+i\vec{k}(\vec{r}-\vec{r}')} G_{ab}(\omega_n, \vec{k})
\label{eq:Green1}
\eea
$(a, b= 1,2)$, where $\omega_n=2\pi nT$ is the $n$th bosonic Matsubara frequency, \be
\sum_{n, \vec{k}}\equiv\sum_{n=-\infty}^{n=\infty}\int d\vec{k}/(2\pi)^3 \nn
\ee
and
\bea
G_{ab}(\omega_n, \vec{k}) = \frac{1}{\omega_n^2 + E_k^2}\begin{bmatrix}
	\epsilon_k+X_2 & \omega_n \\
	-\omega_n & \epsilon_k+X_1
	
\end{bmatrix}.
\label{eq:Gab}
\eea
In Eq.\,\eqref{eq:Gab} $E_k$ corresponds to the dispersion of quasi-particles (Bogolons)
\bea
E_k = \sqrt{\epsilon_k+X_1}\sqrt{\epsilon_k+X_2}.
\label{eq:energy}
\eea
where the self-energies $X_1$ and $X_2$ which are given by
\begin{subequations}
	\begin{align}
	X_1 =\Sigma_{n}+\Sigma_{an}-\mu \label{eq: x1}\\
	X_2 = \Sigma_{n}-\Sigma_{an}-\mu
	\label{eq: x2}
	\end{align}
\end{subequations}
may be considered as variational parameters instead of $\Sigma_{n}$, $\Sigma_{an}$ (see below).  From
Eq.\,(\ref{eq:energy}), it is clear that if one of $X_1$ or $X_2$ is negative then the excitation energy $E_k$  becomes imaginary and the system becomes dynamically unstable in the sense that infinitesimal perturbations will grow exponentially with time. This dynamic instability may start at low momenta and destroy the whole BEC. The stability condition in the equilibrium requires, $X_1 \geq 0$, $X_2 \geq 0$ as well as
\bea
\frac{\partial \Omega}{\partial X_1}=0, \quad  \frac{\partial \Omega}{\partial X_2}= 0.
\label{eq: partial}
\eea
The parameter $\delta$ should be set equal to unity, $\delta=1$, at the end of the calculations \cite{zamos}.\\
(4) After subtraction of discontinuous and one particle reducible diagrams, one obtains the free energy $\Omega$ as a function of $\chi, \bar{\chi}, X_1, X_2$, i.e., $\Omega(\chi, \bar{\chi}, X_1, X_2)$, where $X_1$ and $X_2$ will be fixed by Eqs.\,(\ref{eq: partial}).

Thus, limiting ourselves to the first order in $\delta$ we obtain (see Appendix A for details) the following expression for
$\Omega$ including EA and DM interactions up to quadratic order in $ \gamma'^2$
\begin{subequations} \label{eq:omega}
	\begin{equation}
	\Omega= \Omega_{SYM} + \Omega_{EA} + \Omega_{DM}.
	\tag{\ref{eq:omega}}
	\end{equation}
	\begin{align}
	\Omega_{SYM}= -\mu \rho_0 + \frac{U\rho_0^2}{2} +\frac{1}{2}\sum_k(E_k-\epsilon_k) +T \sum_k \ln (1-e^{-\beta E_k})&\nn
	\qquad\qquad\qquad\qquad + \frac{1}{2}(\beta_1 B +\beta_2 A) +\frac{U}{8}(3A^2+3B^2+2AB), \label{eq:omega1}\\
	\Omega_{EA}= \frac{\gamma \rho_0}{2}(\xi^2+ \bar{\xi}^2) +\frac{\gamma}{2}(B-A),& \label{eq:omega2}	\\
	\Omega_{DM}  = -i\gamma' (\bar{\xi}-\xi) \sqrt{\rho_0} -\frac{\gamma'^2}{X_2},& \label{eq:omega3}
	\end{align}
\end{subequations}
where
\begin{subequations}
	\begin{align}
	\beta_1& =-\mu -X_1 + \frac{U\rho_0}{2} (\xi^2+ \bar{\xi}^2 + 4)\\
	\beta_2& = - \mu -X_2 - \frac{U\rho_0}{2}(\bar{\xi}^2 +\xi^2 - 4)\\
	A&= T\sum_{k,n} \frac{\epsilon_k+X_1}{\omega_n^2+E_k^2}= \sum_k W_k \frac{\epsilon_k+X_1}{E_k}\\
	B&= T\sum_{k,n} \frac{\epsilon_k+X_2}{\omega_n^2+E_k^2}= \sum_k W_k \frac{\epsilon_k+X_2}{E_k}
	\end{align}
\end{subequations}
and $\xi=e^{i\Theta}$, $\bar{\xi}=e^{-i\Theta}$ with $\Theta$ the phase angle of the condensate, $W_k = (1/2)\coth(\beta E_k/2)=1/2 +f_B(E_k)$, $f_B (x)=1/(e^{\beta x}-1)$.

In contrast to Ref.\,\cite{Sirker2} present approximation includes DM interaction up to the second order
(the last term in Eq.\,\eqref{eq:omega3}) and takes into account the anomalous density $\sigma$.
For a homogeneous system the normal and anomalous densities are defined as $\rho_1 = \int \langle\ti{\psi^+}(r)\ti{\psi}(r)\rangle d\vec{r}$ and $\sigma=\int d\vec{r}\langle\ti{\psi}(r)\ti{\psi}(r)\rangle$ respectively, and may be calculated using the Green functions given in Eqs.\,\eqref{eq:Green1} and \eqref{eq:Gab}. As a  result they have the following explicit form
\begin{subequations}
	\begin{align}
	\rho_1 &= \frac{A+B}{2} = \sum_k\left[\frac{W_k(\epsilon_k+X_1/2 +X_2/2)}{E_k} -\frac{1}{2} \right] \equiv\sum_k \rho_{1k} \label{eq:17a}\\
	\sigma &= \frac{B-A}{2} =\frac{(X_2-X_1)}{2} \sum_k \frac{W_k}{E_k}  \equiv\sum_k \sigma_{k}
	\label{eq:17b}
	\end{align}
\end{subequations}
The total density of triplons per dimer is the sum of condensed and uncondensed fractions:
\bea
\rho =\frac{N}{V}=\rho_0 +\rho_1
\eea
which defines the uniform magnetization per dimer $M=g\mu_B\rho$. The variational parameters $X_1$, $X_2$ satisfy the minimization conditions (\ref{eq: partial}) and may be calculated as the positive solutions of the following algebraic equations
\begin{subequations}
	\begin{align}
	X_1&= 2U\rho + U\sigma -\mu + \frac{U\rho_0(\xi^2+ \bar{\xi}^2)}{2}+\gamma +\frac{2\gamma'^2 D_1}{X_2^2} \label{eq:X1}\\
	X_2&=2U\rho - U\sigma -\mu - \frac{U\rho_0(\xi^2+ \bar{\xi}^2)}{2}-\gamma -\frac{2\gamma'^2 D_2}{X_2^2}
	\label{eq:X2}
	\end{align}
\end{subequations}
where
\begin{subequations}
	\begin{align}
	A_1'&= \frac{\partial A}{\partial X_1}= \frac{1}{8}\sum_{k}\frac{(E_k W_k' + 4W_k)}{E_k}\\
	A_2'&= \frac{\partial A}{\partial X_2}=\frac{1}{8}\sum_{k}\frac{(\epsilon_k + X_1)^2(E_k W_k' - 4W_k)}{E_k^3}\\
	B_1'&= \frac{\partial B}{\partial X_1} = \frac{1}{8}\sum_{k}\frac{(\epsilon_k + X_2)^2(E_k W_k' - 4W_k)}{E_k^3}\\
	D_1&=\frac{A_1'}{\bar{D}}; \quad D_2=\frac{B_1'}{\bar{D}}; \quad \bar{D}=A_1'^2 - A_2'B_1'\\
	W_k'& =\beta (1-4W_k^2)=\frac{-\beta}{\sinh^2(\beta E_k/2)}.
	\label{eq:parameter}
	\end{align}
\end{subequations}
Now we are in the position of proving the relations (\ref{eq: x1}) and (\ref{eq: x2}). Using Eq.\,(\ref{eq:Gab}) and setting $U=\gamma=\gamma'=0$ in Eqs.\,(19) one obtains
\begin{subequations}
	\begin{align}
	G^{-1}(\omega_n, \vec{k})&= \begin{bmatrix}
	\epsilon_k+X_1 & -\omega_n \\
	\omega_n & \epsilon_k+X_2
	\end{bmatrix},
	\label{eq:G1}\\
	(G^{0}(\omega_n, \vec{k}))^{-1}&= \begin{bmatrix}
	\epsilon_k-\mu & -\omega_n \\
	\omega_n & \epsilon_k-\mu
	\end{bmatrix}.
	\label{eq:G0}
	\end{align}
\end{subequations}
Then Eqs.(9) lead to
\bea
\Sigma_{n} =\mu +\frac{X_1+X_2}{2}, \quad \Sigma_{an} =\frac{X_1-X_2}{2}.
\label{eq:10}
\eea
This presents Hugenholtz-Pines (HP) theorem {\cite{Enomoto,HP,HM}} for a pure BEC in terms of $X_1, X_2$ as
\bea
\Sigma_{n}-\Sigma_{an}-\mu=X_2=0,
\eea
which reveals the Goldstone mode with a gapless energy dispersion
\bea
E_k =\sqrt{\epsilon_k+X_1} \sqrt{\epsilon_k+X_2}\mid_{X_2=0}=\sqrt{\epsilon_k(\epsilon_k+X_1)} \approx ck + O(k^3),
\label{eq:energyL}
\eea
where $c$ is the sound velocity of the single particle excitation in the low-energy limit at the center of Brillouin zone.

In practice some qualitative preliminary calculations at $T=0$ may be performed on the basis of classical effective potential:
\bea
U_{eff} = -\mu \chi \bar{\chi} +\frac{U\chi^2\bar{\chi}^{2}}{2} + \frac{\gamma (\chi^2 +\bar{\chi}^{2})}{2} -i \gamma' (\bar{\chi} -\chi)
\label{eq: UeffTA}
\eea
which corresponds to the case of neglecting quantum fluctuations in Eq.\,(\ref{eq:psi}) (i.e., $\ti{\psi}=0$) and taking only the appropriate terms in Eqs.\,(\ref{eq:omega}) and {(17)}. For numerical analysis it is convenient to express
Eq.\,(\ref{eq: UeffTA}) in dimensionless form as
\bea
\ti{U}_{eff} =r_{0}^4+2r_{0}^2\left(\ti{\gamma}\xi^2 +\ti{\gamma}\bar{\xi}^2 -2\right) +\frac{4ir_{0}\ti{\gamma}'(\xi - \bar{\xi})}{\sqrt{\rho_{0c}}},
\label{eq: UeffTB}
\eea
where $\ti{U}_{eff}=8U_{eff}U/\mu^2$,
$r_{0}=\sqrt{\rho_0 /\rho_{0c}}$,
$\ti{\gamma}=\gamma/\mu$, $\ti{\gamma}'=\gamma'/\mu$ and $\rho_{0c}=\mu/2U$ is the critical density of pure BEC. In particular, in the isotropic case, when $\gamma=\gamma'=0$, $\ti{U}_{eff}(r_0)$ has its extrema at $r_{0}=0, \pm \sqrt{2}$ corresponding to the symmetric ($r_{0}=0$) and SSB cases ($r_{0}=\pm \sqrt{2})$ and does not depend on the phase angle (see Fig.\,1a). In the next section we shall discuss the phase dependence of ground state of the system by studying minimal points of $U_{eff}$ and $\Omega$ in detail.

\section{The condensate fraction and its phase}

The condensate fraction $\rho_0$ and its phase $\xi =\exp(i\Theta)$ may be found from Eqs.\,(5) and (\ref{eq:omega}) as

\begin{subequations}
	\begin{align}
	\frac{\partial\Omega}{\partial\rho_0}=\cos2\Theta (U\sigma + \gamma) + U (\rho_0 +2\rho_1)-\mu -\frac{\gamma' \sin\Theta}{\sqrt{\rho_0}} =0,   \label{eq:cos1}\\
\frac{\partial^2\Omega}{\partial\rho_{0}^{2}}=U+\frac{\gamma' \sin \Theta}{2\rho_{0}^{3/2}}\geq 0,
\label{drho02}\\
	\frac{\partial\Omega}{\partial\Theta}=2\cos\Theta \left(2\rho_0 (U\sigma + \gamma) \sin\Theta +\gamma'\sqrt{\rho_0}\right) =0,
	\label{eq:cos2}\\
	\frac{\partial^2\Omega}{\partial\Theta^{2}}=-2\rho_0 \cos2\Theta (U\sigma + \gamma) + \sin\Theta \gamma' \sqrt{\rho_0}\geq 0.
	\label{eq:cos3}
	\end{align}
\end{subequations}

The first couple of the equations is convenient to determine $\rho_0$, while the other two are used to fix the phase angle. In particular, Eq.\,(\ref{eq:cos2}) has two branches of solutions for the phase angle which we call mode-1 and
mode-2 as
 \bea
\Theta=  \left\{
 \begin{array}{ll}
 	 -\arcsin (\tilde{s}) +2\pi n,  & \xi=\sqrt
 {
 1- {\tilde{s}}^2
 }
 -i\tilde{s}: \hbox{mode-1}\\
  \frac{\pi}{2}+ \pi n,  & \xi=\pm i  : \hbox{mode-2}\\
 \end{array}
 \right.
 \label{eq:theta}
 \eea
where $\tilde{s}={\gamma'}/{2\sqrt{\rho_0}(U\sigma+\gamma)}$, and $n=0, \pm 1,\pm 2 \ldots$ may be interpreted as a topological number. The question arises which mode of the
phase could be realized in nature?
 To answer this question we discuss particular cases separately, rewriting Eqs.\,(19) and (27) in following equivalent form:
\begin{subequations}
	\begin{align}
	X_1&=U\sigma + U\rho_0 + \gamma + \frac{(\xi^2+\bar{\xi}^2)(U\rho_0-\gamma-U\sigma)}{2}-\frac{i\gamma'(\xi - \bar{\xi})}{2 \sqrt{\rho_0}} + \frac{2\gamma'^2 D_1}{X_2^2} \label{eq:X1'}\\
	X_2&= -U\sigma + U\rho_0 - \gamma - \frac{(\xi^2+\bar{\xi}^2)(U\rho_0+\gamma+U\sigma)}{2}-\frac{i\gamma'(\xi - \bar{\xi})}{2 \sqrt{\rho_0}} - \frac{2\gamma'^2 D_2}{X_2^2} \label{eq:X2'}\\
	\mu& = U\rho_0 + 2U\rho_1 + \frac{(\xi^2+\bar{\xi}^2)(U\sigma + \gamma)}{2} + \frac{i\gamma'(\xi - \bar{\xi})}{2 \sqrt{\rho_0}} \label{eq:mu'}\\
	\frac{\partial\Omega}{\partial\Theta}& = (U\sigma + \gamma) (\xi^2-\bar{\xi}^2)\sqrt{\rho_0} + i\gamma'(\xi+ \bar{\xi})=0 \label{eq:O}\\
	\frac{\partial^2\Omega}{\partial \Theta^2}&= - 2\rho_0(\xi^2+\bar{\xi}^2)(U\sigma + \gamma) - i (\xi - \bar{\xi})\gamma' \sqrt{\rho_0}\geq 0 \label{eq:PO}
	\end{align}
\end{subequations}

\subsection{Ideal Bose gas: $\gamma=\gamma'=U=0$}
It is easily understood that in this case $\partial \Omega/\partial \Theta$ in Eq.\,(\ref{eq:O}) equals exactly to zero. This means that the phase of BEC of an ideal gas may be arbitrary.

\subsection{Isotropic case : $\gamma=\gamma'=0, U\neq 0$}
\label{Isotropic}
First, we note that in the HFP approximation with $\sigma=0$, Eq.\,(\ref{eq:O}) is satisfied for any finite $\xi$. Thus, this version of MFA allows for BEC to have any arbitrary phase angle. However, inclusion of anomalous density into the picture makes the choice for $\xi$ more restricted.
Indeed, for $\sigma\neq 0$, from Eq.\,(\ref{eq:O}), i.e., $U\sigma  (\xi^2-\bar{\xi}^2)=0$, we observe that pure BEC may possess a purely complex, $\xi=\pm i$ (mode-2) or real $\xi=1$ (mode-1) phases. The effective potential (\ref{eq: UeffTA}) at zero temperature is
\bea
U_{eff}|_{\gamma=0,\gamma'=0}=-\mu\rho_0 +\frac{U\rho_0^2}{2},
\eea
and hence its minimum corresponding to the ground state does not depend on the phase, as it is ordinary for a
system with SSB (see Figs.\,1a and 1d). Therefore, one may expect that both modes with $\xi =\pm i$ and $\xi=\pm 1$ are equivalent.\footnote{Since in this case $\xi$ and $\bar{\xi}$ appear in $\Omega$ in the second order we may
	omit the sign ($\pm$).} However, it can be shown that $\xi=i$ case leads to a dynamic instability when quantum fluctuations are taken into account.

\begin{figure}[htbp!]
	\includegraphics[angle=0,width=12cm]{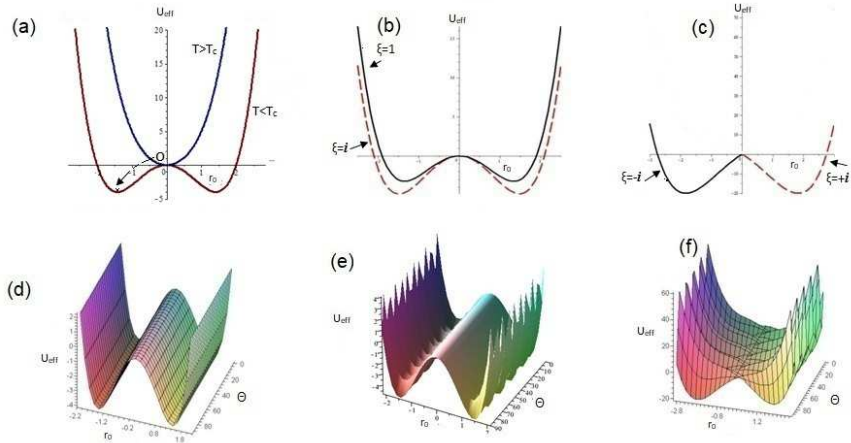}
	\caption{$2D $  and $3D$ effective potentials in units of $8U/\mu^2$ versus the dimensionless order parameter
		$r_0=\pm \sqrt{\rho_0/\rho_{c0}}$, for various cases of interaction: (a, d) $\gamma=\gamma'=0$; (b, e) $\gamma \neq 0, \gamma'=0$; and (c, f) $\gamma=0, \gamma' \neq0$.
		Fig.\,1a illustrates the SSB mechanism. During the  phase transition at $T\sim T_c$
		a weak stray external force gives a small kick, making the particle to move
		``down", say to the left. The system then amplifies this small asymmetry simulating
		a nucleation \cite{snoke}.
	}
	\label{fig:U}
\end{figure}

In fact, when $\xi= i$ the main equations Eqs.\,(\ref{eq:X1}) and (\ref{eq:X2}) are simplified as
\begin{subequations}
	\begin{align}
	X_1&= 2U\rho - U\rho_0 + U\sigma -\mu  \label{eq: N X1}\\
	X_2&=2U\rho + U\rho_0 - U\sigma -\mu
	\label{eq: N X2}
	\end{align}
\end{subequations}
and Eqs.\,(\ref{eq:cos2}) and (\ref{eq:mu'}) with $\Theta=\pi/2$ may have a solution $\rho_0(T=T_c)=0$ i.e., pure BEC phase transition at $T\leq T_c$ occurs. Thus, in the condensate phase Hugenholtz-Pines theorem \cite{HP} should work,
\footnote{Note that, in derivation of this relation for finite temperature, Hoherberg and Martin  did not assume any specific value for the phase $\xi$    {\cite{HM}}.}
 which in our notation means just $X_2=0 $ (see Eq.(\ref{eq: x2})). Now, excluding $\mu$ from Eq.\,(\ref{eq: N X2}) and inserting it into Eq.\,(\ref{eq: N X1})  one obtains
\begin{subequations}
	\begin{align}
	X_1&= -2U\rho_0 + 2U\sigma \label{eq: sigma1}\\
	X_2&=0,
	\end{align}
\end{subequations}
with
\be
\sigma = -\frac{X_1}{2} \sum_{k} \frac{1}{E_k}\left(\frac{1}{2} + \frac{1}{e^{\beta E_k}-1}\right).
\label{eq: sigma2}
\ee
Stability condition requires that $X_1$ and $X_2$, in dispersion $E_k=\sqrt{\epsilon_k+X_1}\sqrt{\epsilon_k+X_2}$ should not be negative, $X_1\geq 0$, $X_2\geq 0$. However, from Eqs.\,(\ref{eq: sigma1}) and (\ref{eq: sigma2}) one can see that $\sigma<0$ and, hence $X_1\leq 0$. Moreover, as seen from Eq.\,(\ref{eq:PO}) the case with $\sigma<0$, $\xi=i$ corresponds not to the minimum, but to the maximum of $\Omega$ . Thus, the case with $\xi=\pm i$, i.e., with $\Theta=\pi/2$  should be excluded for pure BEC.

On the other hand, the mode-1 with $\xi=1$, $\Theta=0$, in Eq.\,(19) with $X_2=0$ leads to
\bea
X_1 =2U(\rho_0 +\sigma)  \label{eq: 26}\\
\mu=2U \rho_1+U \rho_0 -U \sigma
\label{eq: 27}
\eea
which has positive solutions at any temperature \cite{our mce}, since the condensate fraction dominates over the anomalous density.

In actual calculations one has to  introduce an additional Lagrange multiplier
$\mu_0$  in accordance with the Yukalov prescription \cite{yukkl} to avoid Hohenberg-Martin dilemma.
The essence of this dilemma in our context is the following. The chemical potential $\mu$ satisfying the Eq.\,(\ref{eq: 27}) does not coincide with the one implicit in Eq.\, (27a). In other words,
in the isotropic case even with proper phase, the Eqs.\,(27a) and (\ref{eq: 27}) can not be satisfied simultaneously
with the same $\mu$. Therefore, it is assumed that
the condensed fraction of particles may have its own chemical potential, $\mu_0\neq \mu$.
Starting with the action
\be
A_{ISO}[\psi,\psi^+]=
	 \int_{0}^{\beta}d{\tau} d\vec{r} \left\lbrace \psi^+ \left[\frac{\partial}{\partial\tau}-\hat{K}\right] \psi -\mu_0\rho_0-\mu\tilde{\psi}\tilde{\psi}^{+}
+ \frac{U}{2}\left( \psi^+\psi\right)^2  \right\rbrace.
	\label{AISO}
\ee
in Eq. (7a) one obtains
\be
 \frac{\partial\Omega}{\partial \rho_0}=U\cos2\Theta+U(\rho_0+2\rho_1)-\mu_0=0\, ,
\label{rho0iso}
\ee
which is used this time to determine not $\rho_0$ but $\mu_0$.
As to the equations (34) and  (35), they may be considered as a system of two coupled equations
with respect to the variational parameter $X_1$ and condensed fraction $\rho_0$, whereas
$X_2$ is set to zero. On the other hand, in the presence of an anisotropy, which breaks  $U(1)$ symmetry explicitly
, there is no Goldstone mode, and hence the requirement as $X_1=0$ or $X_2=0$  is not needed. This excludes the necessity of   introduction
of an additional chemical potential $\mu_0$, since there would be no  contradiction like Hohenberg-Martin dilemma \cite{23 our aniz}.

 Therefore we can conclude that the phase of pure BEC with SSB in equilibrium can be realized only when $\xi =\pm 1$, that is $\psi =\pm \sqrt{\rho_0}+\ti{\psi}$, otherwise Hugenholtz-Pines theorem will not be satisfied or the system will possess an instability. It is interesting to note that deriving this conclusion we have not referred to a field induced BEC. This is why
Eqs.\,(\ref{eq: 26}) and (\ref{eq: 27}) are exactly the same as it was obtained for uniform atomic gases \cite{yukkl}. Therefore, we may argue that the phase angle of any uniform BEC originating from SSB should be equal to $\Theta=\pi n$.
There is every likelihood that when condensates start to form, e.g., in a rectangle trap \cite{rect_exper,rect_science,rect_yak}, those BEC with other phases leave the condensate, such that only those formations with  with $\Theta=\pi n$ survive. The fact that even in the case of spatially inhomogeneous trapped Bose gases condensate has a fixed phase was experimentally observed many years ago \cite{andrews}. Such experiments seem to support that, as a result of SSB the order parameter exists with a fixed real phase, not only in homogeneous infinite volume system, but also in inhomogeneous system of finite size.

At the end of this subsection, we present an interesting prediction. It concerns the sign of $\sigma$ the anomalous density, $\sigma=(\langle \ti{\psi} \ti{\psi} \rangle+\langle \ti{\psi}^+ \ti{\psi}^+\rangle)/2$. On the one hand, some authors \cite{andersen, griffith, bodjuma} predict $\sigma<0$ especially at low temperatures. On the other hand, using dimensional regularization {\cite{yukkl,ouryee}} gives $\sigma(T\approx 0)>0$. We predict that it should be negative at least for pure BEC in the full range of temperatures, $T\leq T_c$ for a finite system when the momentum integration is performed
in a finite volume. In fact, for $\xi =1$, $ \gamma'=\gamma=0$ case, the condition of minimum of
$\Omega$ with respect to the phase Eq.\,(\ref{eq:PO}) simplifies as
\bea
-U \sigma \rho_0 \geq 0,
\eea
which requires $\sigma \leq 0$ at $T< T_c$ for a repulsive interparticle interaction. As to the normal phase, $T>T_c$,
$\rho_0 (T>T_c)=0$ and $\sigma(T>T_c)=0$ as expected. Thus, for a pure BEC $\sigma \leq 0$, otherwise the state with phase $\chi = \pm \sqrt{\rho_0}$ will be a maximum, not a minimum of $\Omega$. The sign of $\sigma$ also controls the sign of
$\Sigma_{an}$.
From Eqs.\,(\ref{eq:10}) and (\ref{eq:17b}) one finds
\bea
\sigma =-\Sigma_{an}   \sum_{k} \frac{1}{E_k}\left(\frac{1}{2} + \frac{1}{e^{\beta E_k}-1} \right) \equiv -\Sigma_{an} R_\sigma.
\eea
Since momentum summation results in a positive value $R_\sigma>0$, $\Sigma_{an}$ is also positive.
Finally, we remark that in Bogoliubov or  HFP versions of MFA the  anomalous density is usually neglected. As it is seen from Eq.\,(\ref{eq:cos2}) for pure BEC, $\gamma=\gamma'=0$, with $\sigma=0$, $(\partial \Omega / \partial \Theta)=0$ for any $\Theta$. However, taking into account the anomalous density locks the phase to $\xi=\pm 1$, making
$\Theta=\pi n$ the only possible outcome.

\subsection{No DM interaction: $\gamma'=0$, but $\gamma\neq0$, $U\neq 0$}
\label{C2}

Here the symmetry is explicitly broken (see Figs.\,1b and 1e) and again, in principle, two modes of the phase are possible: $\xi=1$ and $\xi=i$. Using Eq.\,(\ref{eq: UeffTA}) the effective potential at $T=0$ may be represented as
\bea
U_{eff}=  \left\{
\begin{array}{ll}
	-\mu\rho_0 +\frac{U\rho_0^2}{2}-\gamma \rho_0,  & \xi= i  \\
	-\mu\rho_0 +\frac{U\rho_0^2}{2}+\gamma \rho_0,  & \xi=1 . \\
\end{array}
\right.
\label{eq:Ueff}
\eea
At the minimum point, $\rho_0=(\mu\pm \gamma)/U$ the potential energy becomes\\
\begin{subequations}
	\begin{align}
	U_{eff}^{min}= -\frac{(\mu+\gamma)^2}{2U},  \quad \xi= i  \label{eq:UminiA}\\
	U_{eff}^{min}= -\frac{(\mu-\gamma)^2}{2U}, \quad \xi=1  \label{eq:UminiB}
	\end{align}
\end{subequations}
The case $\xi=i$ has been studied by Delamore et al. \cite{Delamore}. Considering the minimum of the potential given in Eq.\,(\ref{eq:UminiA}) the authors came to the following conclusion. A field induced BEC and its host crystal that lowers the total energy, have a tendency to increase or even create exchange anisotropy perpendicular to the exchange magnetic field above $H_c$. That is the more intensive $\gamma$, the deeper the ground state. On the other hand, as seen from
Eq.\,(\ref{eq:UminiB}) the case $\xi=1$ leads to the opposite conclusion: minimum of the effective potential ``moves up" with increasing $\gamma<\mu$. Moreover, in favor of the complex case $\xi=i, \chi=i\sqrt{\rho_0}$ implies that the ground state lies lower than the case $\chi=\sqrt{\rho_0}$; $U_{eff}^{min}(\xi=i)<U_{eff}^{min}(\xi=1)$ as illustrated in
Figs.\,1b and 1e.
In other words, the system prefers to decrease $\gamma$. Below we shall show that a more detailed analysis including quantum fluctuations require to exclude $\xi=i$ phase.

In fact, letting $\xi=i$ the main Eqs.(29) become
\begin{subequations}
	\begin{align}
	X_1&=2(U\sigma+\gamma),  \label{eq:X1''}\\
	X_2&=2U\rho_0, \label{eq:X2''}\\
	\mu&=U\rho_0 +2U\rho_1 -U\sigma -\gamma.
	\label{eq:mu''}
	\end{align}
\end{subequations}
Due to the presence of exchange anisotropy ($\gamma\neq0$) there is no SSB and hence we can not exploit HP theorem as it was done in the previous subsection.
On the other hand, from Eq.\,(\ref{eq:mu''}) one notes that it has $\rho_0=0$ solution, which corresponds to a critical temperature $T_c$, above which there is a normal phase with $\rho_0(T>T_c)=0$, $\sigma(T>T_c)\approx 0$. This means that $X_1(T>T_c) \approx 2\gamma$ is constant and $X_2(T>T_c)=0$. Thus, the dispersion with these self-energies
\bea
E_k(T>T_c) = \sqrt{(\epsilon_k+X_1)(\epsilon_k+X_2)}\approx\sqrt{\epsilon_k}\sqrt{\epsilon_k+2\gamma} =ck + O(k^3).
\label{eq:energyG}
\eea
has a linear dependence at small momentum. In other words, the energy dispersion becomes gapless at high temperatures. This is in contradiction with experiments as well as common sense. Thus, taking into account the quantum fluctuations, we have shown that using the complex phase of BEC, $\xi=i$, leads to an unphysical result in presence of EA interaction. As to the case with $\xi=1$ it leads to an expected physical prediction:
$X_1(\xi=1, T>T_c)\approx X_2(\xi=1, T>T_c)\approx 2U\rho-\mu\equiv-\mu_{eff}$, $E_k\approx\epsilon_k-\mu_{eff}$ \cite{23 our aniz}.

The stability condition ${\partial^2\Omega}/{\partial \Theta^2}\geq 0$ sets an upper bound to the intensity of EA interaction too. In fact, this time Eq.\,(\ref{eq:PO}) with the real phase $\xi =1$ reduces to
\bea
\rho_0(U\sigma +\gamma)\leq 0.
\label{eq:35}
\eea
At high temperatures, this is satisfied due to $\rho_0(T>T_c)=0$. However, at small temperatures this inequality demands  $\gamma\leq U |{\sigma}|$ at any temperature, $T\leq T_c$. For instance,
for the triplon gas in TlCuCl$_3$ the anomalous density is of order $\sigma|_{T=0}\approx -0.5 \times 10^{-2}$ and $U\approx 300$\,K \cite{our mce}, so $\gamma\leq 1.5$\,K.
Eq.\,(\ref{eq:35}) answers affirmatively the question if one should take into account anomalous density for crossover transition, when the symmetry is explicitly broken. Otherwise one deals with $\rho_0 \gamma \leq 0$ which cannot be satisfied with positive ($\gamma>0$) EA interaction. Note that the importance of $\sigma$ was discussed also in
Refs.\,\cite{ourANN,bodjuma,yukalov2005}.

\subsection{EA and DM interactions ($\gamma\neq0, \gamma'\neq0$) }
\label{C3}
We now explore the combined effect of both anisotropies to the phase and condensate fraction of triplon BEC. In the previous subsection we have shown that the EA retains the phase and the nature of BEC unchanged. In the present subsection we show that the presence of DM interaction changes the picture dramatically.

As seen from Eq.\,(\ref{eq:theta}), in this case three variants for $\xi$ are possible: $\xi=+i$, $\xi=-i$ and $\xi=\exp(i\Theta)$ with $\sin\Theta=-\gamma'/ ({2\sqrt{\rho_0}(U\sigma +\gamma)})$. The
corresponding effective potential is illustrated in Figs.\,1c and 1f. Moreover, due to the last term in Eq.\,(\ref{eq:mu'}) the condensate fraction remains finite at any finite temperature. For this reason the critical temperature $T_c$ may be defined from the minimum of $M$ as $(\partial\rho/\partial T)_{T=T_c}=0$.

First, we discuss the most interesting choice with
\bea
\sin\Theta=-\frac{\gamma'} {{2\sqrt{\rho_0}(U\sigma + \gamma)}}
\label{eq:sinT}
\eea
This phase angle is rather attractive for the following reasons.
(i) It is temperature dependent.
(ii) One can decrease $\gamma'$ smoothly up to $\gamma' = 0$ obtaining $\sin\Theta=0$, which corresponds to the pure BEC case with the real phase $\xi=1$, as has been discussed before.
(iii) Since for the triplon system $\Theta$ corresponds to the angle between $y$-axis and staggered magnetization $M_{\perp}$ (see Fig.\,\ref{fig3}a {in Sect. V}), one may observe from Eq.\,(\ref{eq:sinT}) that changing the temperature leads to a change in $\sigma$ and $\rho_0$, hence modifies the direction of $M_{\perp}$ in the (xy)-plane. In other words, it seems that, by measuring the direction of $M_{\perp}$, one will be able to get information about the phase of triplon condensate. However, as we show below this phase is unphysical.

Inserting Eq.\,(\ref{eq:sinT}) with $\xi=e^{i\Theta}$ into Eqs.\,(\ref{eq:X1'}) and (\ref{eq:X2'}) gives
\begin{subequations}
	\begin{align}
	X_1=2U\rho_0-\frac{U\gamma'^2}{2(U\sigma +\gamma)^2} + \frac{2\gamma'^2D_1}{X_2^2} \label{eq:36a}\\
	X_2=-2(U\sigma +\gamma)+\frac{U\gamma'^2}{2(U\sigma +\gamma)^2} - \frac{2\gamma'^2D_2}{X_2^2}.
	\label{eq:36b}
	\end{align}
\end{subequations}
It is easily understood that at high temperatures $(T\geq T_c)$ $X_1\approx X_2$ and both are rather large $(X_ 2(T \gg T_c)\gg X_2(T=0))$ \footnote{At high temperatures, the anomalous density given by Eq.\,(\ref{eq:17b}) vanishes, and the energy $E_k \approx (\epsilon_k +X_2)$ is naturally expected to increase.}. Therefore, in this limit Eq.\,(\ref{eq:36a}) leads to
\bea
X_1(T\gg T_c) \approx -\frac{U\gamma'^2}{2\gamma^2}\leq 0
\label{eq:37}
\eea
which corresponds to a dynamic instability. Moreover, Eq.\,(\ref{eq:PO}) with $\Theta$ given by Eq.\,(\ref{eq:sinT}) is not satisfied, which means that, this solution corresponds to the maximum, not the minimum of $\Omega$.

We now discuss the case $\xi=-i$ and show that this should also be excluded. In fact setting $\xi=-i$ in (\ref{eq:X1'}) and (\ref{eq:X2'}) we obtain
\begin{subequations}
	\begin{align}
	X_1&=2U\sigma-\frac{\gamma'}{\sqrt{\rho_0}} + \frac{2\gamma'^2D_1}{X_2^2} + 2\gamma \label{eq:38a}\\
	X_2&=2U\rho_0-\frac{\gamma'}{\sqrt{\rho_0}} - \frac{2\gamma'^2D_2}{X_2^2}.
	\label{eq:38b}
	\end{align}
\end{subequations}
It is easily understood that, at high temperatures ($\rho_0\rightarrow 0$), the second term in Eqs.\,(\ref{eq:38a}) and (\ref{eq:38b}) will dominate and hence, both of $X_1$ and $ X_2$ become negative. Thus, we come to the conclusion that when
$\gamma'\neq0$ the only phase with $\xi=+i$ is accessible even in the HFP approximation.
The main equations Eqs.\,(\ref{eq:X1'}), (\ref{eq:X2'}) and (\ref{eq:mu'})  for the case $\gamma=0$, $\gamma'\neq0$ with $\xi=+i$ have the form
\begin{subequations}
	\begin{align}
	X_1&=2U\sigma +\frac{\gamma'}{\sqrt{\rho_0}}+\frac{2\gamma'^2 D_1}{X_2^2}  \label{eq:GX1}\\
	X_2&=2U\rho_0 +\frac{\gamma'}{\sqrt{\rho_0}}-\frac{2\gamma'^2 D_2}{X_2^2} \label{eq:GX2}\\
	\mu&=U(\rho_0+2\rho_1-\sigma)-\mu-\frac{\gamma'}{\sqrt{\rho_0}}.  \label{eq:coscos}
	\end{align}
\end{subequations}
We shall discuss the properties of these equations in detail in a separate publication. Here we note that for $\gamma'\neq 0$ Eq.\,(\ref{eq:coscos}) has no solution with $\rho_0=0$, which prevents the existence of a critical
temperature defined as $\rho_0(T=T_c)=0$. Hence there is no an ordinary phase transition from BEC to a normal phase but a crossover.

The results of this  section are summarized in Table I.
Thus we come to the conclusion that, similarly to the spontaneous magnetization phenomena, where at $T<T_{curie}$ the spins align in one direction without application of external magnetic field, a Bose gas at $T<T_c$ acquires a condensate fraction with a certain order parameter, $\rho_0$ and a certain phase, $\xi=e^{i\Theta}$ allowed by the conditions of stability against quantum and thermal fluctuations as illustrated in {Figs.\,1}.

\begin{table}
	\caption{Possible phases and transitions in homogeneous BECs. SBS and EBS correspond to spontaneous and explicit breaking symmetry cases, respectively.}
	\begin{tabular}{|L|L|c|c|L|L|}
		\hline
		BEC type & Interaction parameters & Phase & Phase angle& Symmetry breaking & {Transition BEC $\rightarrow$ normal phase}\\
		\hline
		Ideal gas& $U=0, \gamma=0, \gamma'=0$ & $\xi=e^{i\Theta}$& arbitrary& SBS & II-order\\
		\hline
		Pure BEC & $U\neq0, \gamma=0, \gamma'=0$ & $\xi=\pm1$& $\pi n$& SBS & II-order\\
		\hline
		{Interacting gas with EA} & $U\neq0, \gamma\neq0, \gamma'=0$ & $\xi=\pm1$& $\pi n$& EBS & II-order\\
		\hline
		{Interacting gas with DM anisotropy} & $U\neq0, \gamma=0, \gamma'\neq0$ & $\xi=+i$& $\pi/2 + 2\pi n$& EBS & crossover\\
		\hline
		{Interacting gas with  both EA and DM anisotropies} & $U\neq0, \gamma\neq0, \gamma'\neq0$ & $\xi=+i$& $\pi/2 + 2\pi n$& EBS & crossover\\
		\hline	
	\end{tabular}
	\label{T:results}
\end{table}

\section{Interference of two condensates and Josephson effect  }

The best way of studying the phase of matter experimentally is through measurements with interference patterns or Josephson junctions. The former is usually performed with atomic Bose condensates and the latter with superconductors or, possibly, quantum magnets. Below we discuss  the consequences of our results summarized in the previous section, to each of these effects.

\subsection{Interference between two Bose condensates}

It is well known that the interference effect occurs due to the phase difference of two matter waves. Nearly ten years before observing the first interference effect between two BECs by the MIT group  \cite{andrews}, Anderson \cite{anderson} had raised his famous question ``Do superfluids that have never seen each other have a well-defined relative phase?" Further this nontrivial question has been reformulated in a philosophical way ``Does the BEC phase appears under the effect of SSB when it is formed, or later, when quantum measurement occurs?" There is no unique answer to these questions {\cite{snoke,mullin2}}. In fact, even if the phase preexists, it is doubtfult wether it can be directly measured because of collisions during a ballistic expansion.
Nevertheless, the existing experiments \cite{andrews, stein, vogels, you} seem to support that as a result of SSB the order parameter exists with a fixed phase, not only in homogenous infinite volume system, but also in inhomogenous system of finite size. This conclusion is consistent with our predictions derived in the previous section.

Qualitatively interference picture can be described in the following simple way. Suppose that we have an order parameter that involves two condensate wave functions, with condensate densities $\rho_{0}^{a}$ and $\rho_{0}^{b}$ in momentum states $\vec{k}_{a}$ and $\vec{k}_{b}$. Then, the density of the combined system is \cite{mullin2}
\bea
\rho_0(\vec{r})=|\sqrt{\rho_0^a}e^{i\vec{k}_{a}\cdot\vec{r}}e^{i\Theta_a}+\sqrt{\rho_0^b}e^{i\vec{k}_b\cdot\vec{r}}e^{i\Theta_b}|^{2}
=\rho_{0}^{ab}[1+x\cos(\vec{k}\cdot\vec{r}+\Theta_{ab})] \label{eq:2.1}
\eea
where $\rho_0^{ab}=\rho_0^a+\rho_0^b$, $\vec{k}=\vec{k}_a-\vec{k}_b$, $\Theta_{ab}=\Theta_b-\Theta_a$, $x=2\sqrt{\rho_0^a\rho_0^b}/(\rho_0^a+\rho_0^b)$ and $\Theta_a$, $\Theta_b$ are initial phase angels.
So, we  have an interference pattern with relative phase $\phi=\vec{k}\cdot\vec{r}+(\Theta_b-\Theta_a)$
to discuss particular cases.

{\it Interference of two pure condensates}\\
In this case as it is seen from Table\,\ref{T:results}    $\Theta_b-\Theta_a=\pi(n_b-n_a)\equiv\pi m$, $m=0,\pm1,\pm2,....$.
From Eq.\,(\ref{eq:2.1}) one obtains $\rho_0(\vec{r})=\rho_0^{ab}(1\pm\cos\vec{k}\vec{r})$.
In particular, when both condensates have the same phase $\xi_a=\xi_b=+1$, one obtains the well-known result, e.g.,
$|\vec{k}\vec{r}|=2\pi n$ corresponds to constructive interference, since  $\cos({\pi}/{2}-\pi n)=\sin\pi n$.

{\it BEC with DM interaction}\\
Now suppose that we have two interfering condensates. One (b) is a pure condensate with SSB and the other (a) includes a tiny DM interaction, given by the linear Hamiltonian (\ref{eq:H4}). The initial phases may be as $\Theta_a=\pi/2$ and $\Theta_b=0$, so $\Theta_{ba}=-\pi/2$. Since $\cos(x-\pi/2)=\sin x$,  (\ref{eq:2.1}) gives
\bea
\rho_0(\vec{r})=\rho_0^{ab}(1+x \sin(\vec{k}\vec{r})).
\label{eq:3.1}
\eea
This means that, in contrast to the previous case, the condition $|\vec{k}\vec{r}|=2\pi n$ will correspond not to a constructive, but to a destructive interference.
Thus, the presence of DM interaction in one of condensates dramatically changes the interference picture, demonstrating its sensitivity to the initial phase.

\subsection{Stationary Josephson effect}

It is well known that Josephson effect can take place due to the phase difference $\Delta\Theta=\Theta_1 -\Theta_2$ between two contacting materials (d.c. effect) or due to the difference of chemical potentials $\Delta\mu=\mu_1-\mu_2$ \cite{imanbook}, (a.c. effect). Besides of superconductors, this effect has been observed in superfluid helium \cite{SFHESE}, where $\Delta \mu \neq 0$ is reached by application of a pressure differential, as well as in Bose condensates in a double well potential, by changing the relative condensate population $\Delta N= (N_1 -N_2)/N \sim \Delta \mu$ \cite{BEC jeff}. A natural question arises if Josephson effect is possible between two magnetic insulators, when each of them contains a triplon condensate.
This question has been discussed some years ago by Schilling and Grundmann \cite{Andreas}. They predicted possible occurrence of Josephson effect between two compounds due to the chemical potential difference $\Delta\mu=g\mu_B(H_{c_1}-H_{c_2})$, say Ba$_3$Cr$_2$O$_8$ and Sr$_3$Cr$_2$O$_8$ separated by nonmagnetic
Ba$_3$V$_2$O$_8$. However, the presence of a DM interaction has not been considered. Here we address the effect of DM interaction on the Josephson effect.

First, we note that in the present article we have been discussing only equilibrium systems. Therefore, studying the
a.c. Josephson effect, which is a dynamical effect, is beyond our scope. However, we may consider the d.c. Josephson effect. In the simple case with identical systems ($\Delta\mu=0$) the change in number of triplons due to tunneling is given by
\bea
\frac{\partial \rho_{01}}{\partial t}=2K \sqrt{\rho_{01} \rho_{02}}\sin(\Delta\Theta)
\eea
i.e., the current is simply proportional to the $\sin(\Delta \Theta)$ \cite{Landau}
\bea
J_{dc} \approx \sin(\Theta_1 -\Theta_2).
\eea
Now, from Table\,I one may come to the conclusion that stationary Josephson effect can take place only when one of the materials has no anisotropy, (or only EA), with $\Theta_1=\pi n$ while the other one has DM anisotropy, $\Theta_2=\pi/2 + \pi n$. In all other cases, e.g., in the contact between two materials with no anisotropy: $\Delta\Theta \approx \pi n$ (n=0, 1, 2...), and hence $J_{dc} =\sin{(\pi n)}=0$. We hope, a proper choice of compounds with suitable material parameters will make the observation of Josephson effect possible in various regimes and in particular verify the above conclusion.

\section{Kibble-Zurek mechanism in staggered magnetization}

The Kibble-Zurek mechanism (KZM) predicts the spontaneous formation of topological defects in systems that cross a second-order phase transition with SSB at a finite rate. The mechanism was first proposed by Kibble \cite{kibble} in the context of cosmology to explain how the rapid cooling below a critical temperature induced a cosmological phase transition resulting in the creation domain structures and e.g., baryons from quark-quark plasma. Further, Zurek has extended this paradigm to the condensed matter physics in the context of vortices in the $\lambda$-transition of superfluid $^{4}$He \cite{Zurek}.
This theory leverages the well-established results of the equilibrium theory of criticallity to make immediate predictions for universal scaling behavior in the nonequilibrium dynamics of passage through a continuous transition.

One of the key parameters of KZM is the quench time $\tau_Q$. In this context the term ``quench" refers to varying a thermodynamic parameter in order to drive the system across the critical point of a phase transition and out of equilibrium for a finite time \cite{polkov}. The ratio between $\tau_Q$ and the relaxation time defines the number of topological defects.  The more rapidly the system passes through the critical point, the shorter the correlation length and the more topological defects will form. Their density and size are given by a simple scaling relations \cite{davis}.

Bose-Einstein condensation of any type of particles is an ideal platform for investigating the Kibble-Zurek paradigm. In this case, domains or topological defects may be as the germs, grains, independent condensates or droplets of condensed atoms inside uncondensed surrounding. Judging by quench or creation time one may classify condensates as ``BEC with rapid quench" and ``BEC with slow quench". ``Fast" or ``slow" depends on the characteristic scales of a system. For example, for a typical atomic BEC $\tau_Q\approx5 s$ \cite{rect_science}.
\begin{figure}[htbp!]
	\centering
	\includegraphics[angle=0,width=10cm]{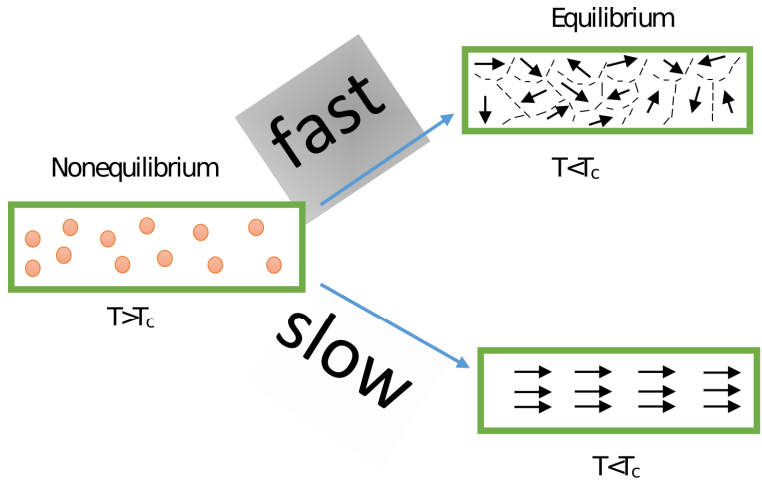}
	\caption{Domain formation during SSB in a homogenous Bose gas. Red points depict thermal atoms. Dashed lines delineate domains over the phase is constant. The arrows independently chosen condensate phase. It is seen that fast quench leads to multiple domain BEC (droplets) while slow quench to a single-domain BEC.}
	\label{fig:1}
\end{figure}
For convenience we present in  Fig.\,\ref{fig:1} a schematic illustration of rapid and slow quenches in gaseous BEC. It is seen that domain walls are positioned randomly and each domain has its own random phase.
Here it should be noted that, these domains, even in the spinor BEC or that of triplons, are not the same as in ordinary ferromagnetic materials. The main difference is that the domains in ferromagnets always exist once created by nature, meanwhile domains under discussion, are born due to SSB and vanish in the normal phase $(T>T_c)$.

There are a number of experimental methods to investigate the KZM. Among them, we can list
(i) the observation of vortex formation during the fast cooling of an atomic gas through the BEC transition \cite{weiler};
(ii) studying the formation of excitations as the quench rate is varied \cite{chen}, which is good for optical lattices;
(iii) directly checking the KZM scaling relations using matter wave interferometer \cite{rect_science};
(iv) identification defects as solitons, which may be observed \cite{lamporese};
(v) studying ``in situ" images of domain walls and spin vortices. The latter was successfully used by Sadler {\it et al.} \cite{sader} to study SSB and KZM in a quenched ferromagnetic spinor BEC by rapidly reducing the magnitude of the applied magnetic field.
Below we propose another method by using spin - gapped quantum magnets.

\begin{figure}[htbp!]
	\begin{minipage}[H]{0.32\textwidth}
		\center{\includegraphics[width=\linewidth]{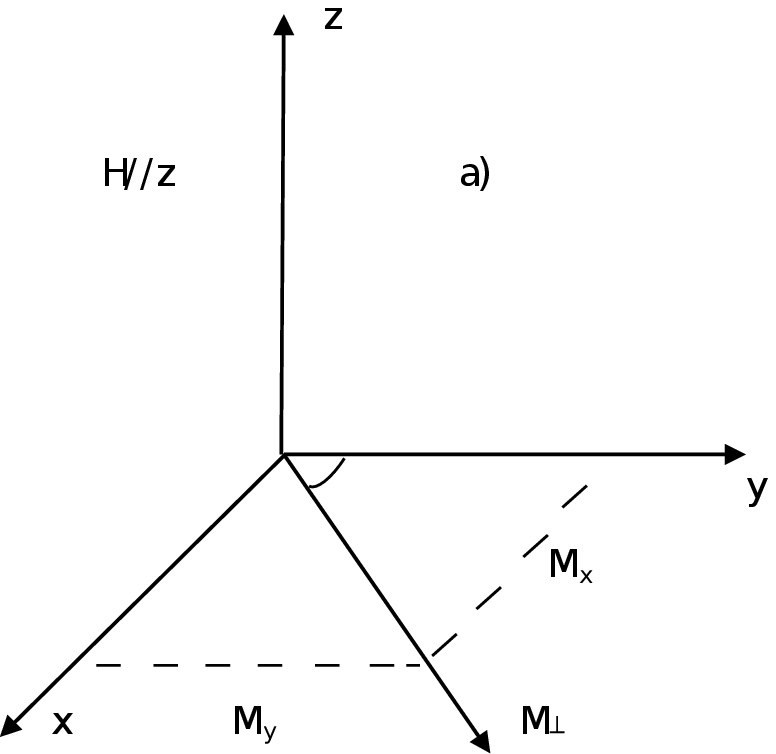}\\}	
	\end{minipage}
	\begin{minipage}[H]{0.32\textwidth}
		\center{\includegraphics[width=\linewidth]{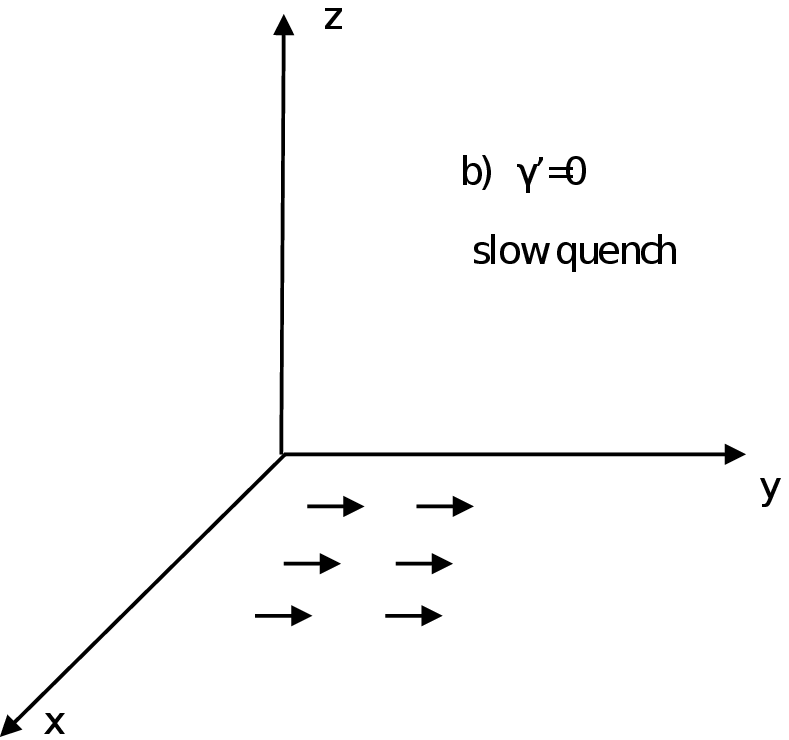}\\}	
	\end{minipage}
	\begin{minipage}[H]{0.32\textwidth}
		\center{\includegraphics[width=\linewidth]{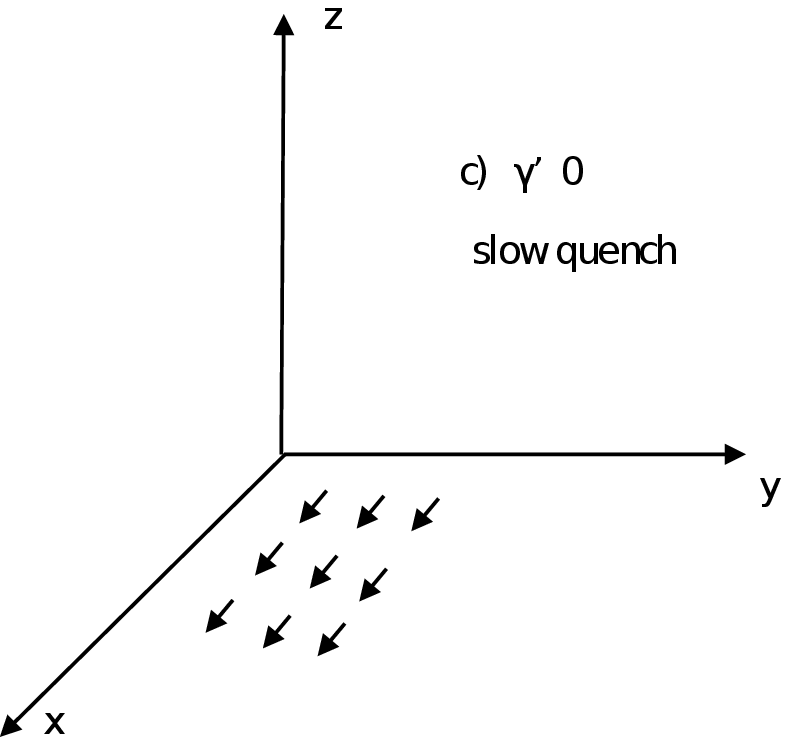}\\}	
	\end{minipage}
	\begin{minipage}[H]{0.32\textwidth}
		\center{\includegraphics[width=\linewidth]{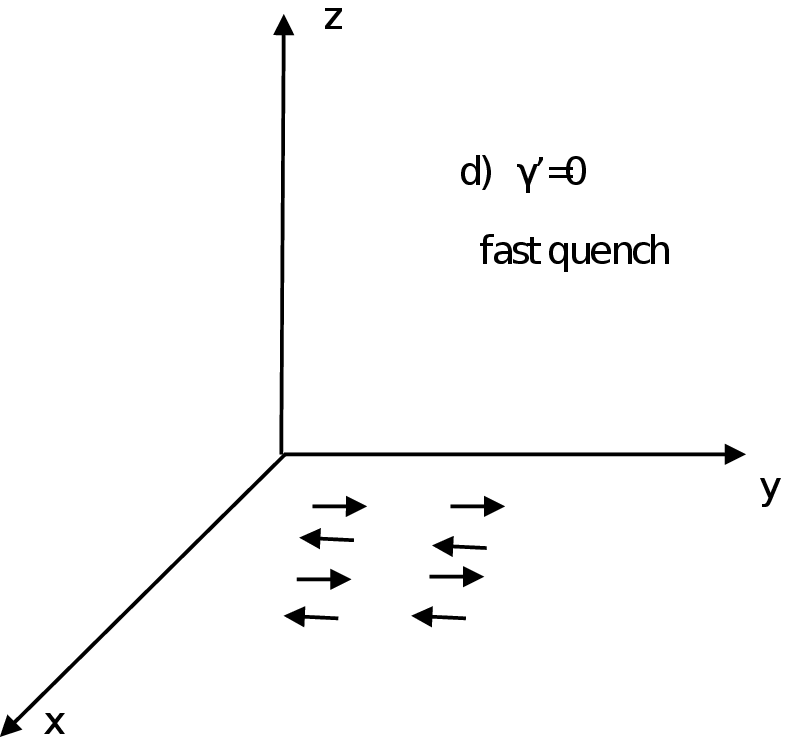}\\}	
	\end{minipage}
	\begin{minipage}[H]{0.32\textwidth}
		\center{\includegraphics[width=\linewidth]{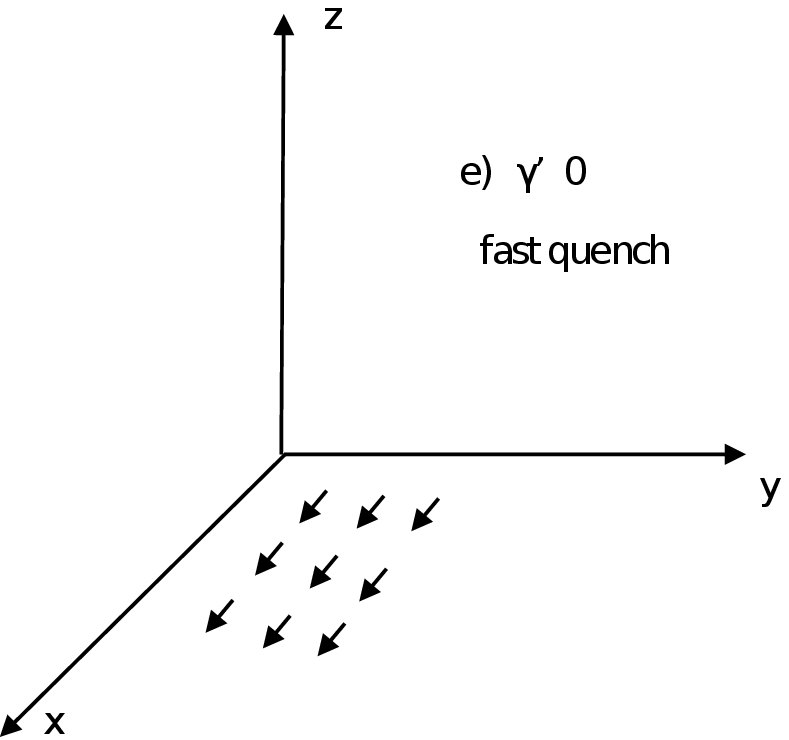}\\}	
	\end{minipage}
	\caption{The vector of staggered magnetization. (a) General representation; (b) and (c) correspond to the case of slow creation of triplon BEC without (b) and with DM anisotropy (c). Figs (d) and (e) are the same as in (b) and (c), respectively, but for fast quench.}
	\label{fig3}
\end{figure}

As it is pointed out in the Introduction, the emergence of a triplon condensate in quantum magnets leads to a finite staggered magnetization $M_\perp$, whose magnitude may be evaluated as $\mid M_\perp\mid=g\mu_B\sqrt{\rho_0/2}$, where  $\rho_0$ is the condensate fraction. As to the direction of the vector $\vec{M}_\perp$, it lies in the $xy$-plane as illustrated in Fig.\,\ref{fig3}a, so $M_\perp=\mid M_x\mid \sin\Theta+\mid M_y\mid \cos\Theta$. Remarkably, the angle
$\Theta$ in this plane corresponds to the phase angle of the condensate wave function \cite{giamarchi}.
Now to study possible Kibble-Zurek mechanism in spin-gapped dimerised quantum magnets we consider ``slow" and ``rapid" quench regimes separately.

\subsection{Slow quench}

In this case we deal with only a single-domain BEC. Thus, from Fig.\,\ref{fig3}b and Table\,\ref{T:results}, one may conclude that $M_\perp$ lies along the $y$-axis ($\Theta=\pi n$) for the system without DM anisotropy, and $M_\perp$ is parallel to the $x$-axis, $\Theta=\pi/2$ when DM interaction is involved, Fig\ref{fig3}c.

\subsection{Fast quench}

This regime can be realized by a rapid quenching of the external magnetic field, which causes the emergence of topological defects, namely triplon condensate droplets. Assume that each droplet is full-grown, and hence can be described by the MFA, as outlined above. Thus, if there is no DM interaction the phase angle of each domain may be equal to one of values such as $\Theta=0,\pi,2\pi... $ and hence $m_\perp=\mid m_y\mid \cos \pi n=\pm\mid m_y\mid$. As a result the total staggered magnetization of the sample $M_\perp=\sum m_\perp^i=0$ (see Fig.\,\ref{fig3}d). On the contrary, in the presence of DM interaction each domain will have a phase angle $\pi/2+2\pi n$, $m_\perp=\mid m_x^i \mid$ which leads to the finite $M_\perp=\sum m_\perp^i=\sum\mid m_x^i\mid\neq 0$ (see Fig.\,\ref{fig3}e). Therefore, we have shown that in a ``fast quench" due to the KZM:

\begin{itemize}
\item [(a)]the total staggered magnetization of a spin-gapped magnet without DM interaction (while EA can be present) vanishes even for $T<T_c$,  $M_\perp(\gamma'=0)=0 $.
\item [(b)]the total staggered magnetization of a quantum magnet with DM anisotropy remains finite, $M_\perp(\gamma'\neq 0)\neq 0$. It will be interesting if this conclusion may be checked experimentally.
\end{itemize}
\section{Conclusion}
We have derived an explicit expression for the grand canonical thermodynamic potential $\Omega$ for a triplon system in dimerized spin-gapped magnets, taking into account both the EA and weak DM anisotropies.  The thermodynamic potential embodies all information about the equilibrium homogenous Bose gas at low temperatures. Particularly, minimization of
$\Omega$ by self-energies $(X_1, X_2)$ yields
two coupled equations, which define
the spectrum of quasiparticles, $E_k=\sqrt{(\varepsilon_k+X_1) (\varepsilon_k+X_2)}$ and the densities of triplons. Minimization by the condensate wave function leads to equations with respect to the phase and condensate fraction of BEC. Satisfaction of these equations, together with stability conditions for BEC, present certain boundaries for the phase angle and condensate fraction of BEC within the framework of MFA, as listed below.
\begin{itemize}
\item [1.]The condensate fraction $\rho_0$, clearly is an intensive parameter depending  on temperature and pressure
while the phase of the condensate is a constant.
\item [2.]The condensate of ideal gas may possess any phase $\xi=e^{i\Theta}$ with an arbitrary real angle $\Theta$. This is true for an interacting gas also, when anomalous density is neglected i.e., $\sigma=0$.
\item [3.] The system without anisotropy, or with only exchange anisotropy ($\gamma\neq0, \gamma'=0$) possesses only real phase with $\xi=\pm 1$ , $\Theta=\pi n$, $n=0,1,2...$ and BEC $\rightarrow$ normal phase transition takes place with a definite critical temperature $T_c$, such that $\rho_0(T\geq T_c)=0$.
\item [4.]The inclusion of DM anisotropy, (at least in the form of a linear Hamiltonian of Eq.\,\eqref{eq:H4}) smears this transition to a crossover, such that $\rho_0(T)$ diminishes asymptotically, and sets the
phase  to $\xi=+ i $, rotating its angle $\Theta$ from 0 to $\pi/2$.
\item [5.]There is a smooth path between BEC($\gamma \neq 0$)$\rightarrow$ BEC($\gamma =0$), in the sense of that by gradually decreasing $\gamma$ one arrives at a pure BEC.
\item [6.]On the other hand, there is no path from a pure BEC to normal phase and crossover transitions: one may slowly decrease $\gamma'$ to get the case with possible $\rho_0(T\geq T_c)=0$, but the phase, which does not explicitly depend on $\gamma'$ remains as $\xi=+ i$, instead of $\xi=\pm1$.
\end{itemize}
Therefore, our mean-field based approach, including the anomalous density predicts that the phase angle $\Theta$ of interacting homogenous BEC may take only discrete values as  $\Theta=\pi n$ or $\Theta=\pi/2+2\pi n$, (n=0,$\pm$1, $\pm$2....) where $n$ can be interpreted as a topological number. This is in contrast to widely used e.g., HFP or  simple
Bogoliubov approximations, where $\Theta$ is allowed to have any arbitrary angle.

Having fixed the problem about the phases we studied the influence of possible phases to the interference and Josephson junction of two Bose condensates. We have shown that when one of the condensates have even a tiny DM interaction the interference picture will change drastically.

Analyzing the simple d.c. Josephson effect between two spin-gapped magnets we have found that there would be no Josephson current when neither of  samples has  DM anisotropy $(\gamma'_{1}=0,\gamma'_{2}=0)$, while the current will be finite, when one of them has a weak DM anisotropy $(\gamma'_{1}=0,\gamma'_{2}\neq 0)$.
We briefly discussed also the consequences of DM interaction to possible Kibble-Zurek mechanism in spin-gapped antiferromagnets. We have shown that ``fast quench" destroys the total staggered magnetization in axially symmetric samples with $\gamma'=0$. On the other hand, in the presence of DM interaction $(\gamma'\neq 0)$, the staggered magnetization remains finite even in the "fast quench" regime. It would be quite interesting to make appropriate experimental measurements to study Kibble-Zurek mechanism in spin-gapped dimerized magnets.

In a following work we shall address how anisotropies modify other physical observables, such as magnetization, heat capacity, etc.\\

{\textbf{Acknowledgment}} \\
AR acknowledges support by TUBITAK-BIDEB, Turkey (2221), LR acknowledges support by
TUBITAK-ARDEB, Turkey (1001), AK is supported by the Ministry of Innovative Development of the
Republic of Uzbekistan, BT is supported by TUBITAK and TUBA, Turkey\footnote{Scientific and Technological Research Council of Turkey}.

\clearpage
\appendix
\section{Effective Hamiltonian in bond-operator representation and   derivation of $\Omega$}
\label{sec:A}
In this appendix we outline the derivation of the Hamiltonian in Eq. (1)
from the basic  spin Hamiltonian and    present the derivation of $\Omega$.

Taking into account the dimerized nature of the ground state, we consider a Heisenberg Hamiltonian given by
\be
{\cal H}= 
\sum_i J_0 {\bfs_{i1}}\cdot {\bfs_{i2}}+\sum_i \sum_{m=1}^{9}J_m {\bfs_{i,1}}\cdot {
\bfs_{{i+\delta {{\bf r_m}}},1}
}
\lab{spinham}
\ee
%
%
where $ i$ denotes locations of dimers. $J_0$ indicates the intradimer spin coupling while
interdimer spin couplings are described by $J_m$ $(m=1,... , 9)$. The two spins within a dimer are labeled by subscripts 1 and 2. Since a dimer is made of two neighboring spins, the interaction between two dimers contains four different spin couplings.
Now following bond - operator representation we introduce following transformations:
\bea
S_{1\alpha} = \frac{1}{2}\left(s^{\dagger}t_\alpha + t^{\dagger}_\alpha s -i\epsilon_{\alpha \beta \gamma}t^{\dagger}_\beta t_\gamma \right),\nn
S_{2\alpha} = \frac{1}{2}\left(-s^{\dagger}t_\alpha - t^{\dagger}_\alpha s -i\epsilon_{\alpha \beta \gamma}t^{\dagger}_\beta t_\gamma \right).
\label{transf}
\eea

with $\alpha=x,y,z$  and  $\epsilon_{\alpha\beta\gamma}$ is the totally antisymmetric unit tensor.The corresponding
auxiliary quasiparticles can, therefore, be called singletons, and triplons, respectiveley.
The restriction on the physical states to be either singlets or triplets leads to the
constraint 
\be
s^{\dagger}s+\sum_\alpha t^{\dagger}_\alpha t_\alpha=1
\lab{chegara}
\ee
Inserting \re{transf} into the spin Hamiltonian \re{spinham} and using constaints \re{chegara} 
one may obtain effective Bose Hamiltonian (1) \cite{Sirker1,dods}.

Now we pass to the derivation of thermodynamic potential $\Omega $ given in Eq.\,(\ref{eq:omega}).
 Inserting Eq.\,(\ref{eq:psi}) into the action Eq.\,(\ref{eq:entropy1}), the latter can be divided into the following parts

\begin{subequations}
	\begin{align}
	S& =S_0 + S_1+S_2+S_3+S_4 \label{eq:S}\\
	S_0&= \int_{0}^{\beta}d\tau\int d\vec{r}\left\lbrace -\mu \rho_0 + \frac{U\rho_0^2}{2} + \frac{\gamma \rho_0}{2} (\xi^2+ \bar{\xi}^2) -i \sqrt{\rho_0} \gamma'(\bar{\xi}-\xi)\right\rbrace \label{eq:S0}\\
	S_1&= \int_{0}^{\beta}d\tau\int d\vec{r} \left\lbrace \left[ i \gamma' + U\rho_0^{3/2}\bar{\xi} +\gamma \xi \sqrt{\rho_0} -\mu \sqrt{\rho_0}\bar{\xi}\right] \ti{\psi} + h.c\right\rbrace \label{eq:S1} \\
	S_2&= \int_{0}^{\beta}d\tau\int d\vec{r} \left\lbrace \ti{\psi}^+\left[ \partial_\tau - \hat{K} + 2U\rho_0 -\mu\right]\ti{\psi} + \frac{\gamma}{2}(\ti{\psi}^2 + \ti{\psi}^{+2}) +\frac{U\rho_0}{2}(\xi^2\ti{\psi}^{+2} + \bar{\xi}^2\ti{\psi}^2) \right\rbrace \label{eq:S2}\\
	S_3&=U \sqrt{\rho_0} \int_{0}^{\beta}d\tau \int d\vec{r} \left\lbrace \bar{\xi} \ti{\psi}^{+} \ti{\psi}^2 + \xi \ti{\psi}^{+2} \ti{\psi} \right\rbrace \label{eq:S3}\\
	S_4&= \frac{U}{2}\int_{0}^{\beta} d\tau \int d\vec{r} \ti{\psi}^+ \ti{\psi}^+ \ti{\psi}\ti{\psi}.
	\label{eq:S4}
	\end{align}
\end{subequations}
Now employing the $\delta$-expansion method, we add to the total action Eq.\,(\ref{eq:entropy1}), the term $ (1-\del)\int_{0}^{\beta} d\tau \int d\vec{r} \left[ \Sigma_{n}(\ti{\psi}^+ \ti{\psi}) + (1/2) \Sigma_{an} (\ti{\psi}^{+}\ti{\psi}^{+} +\ti{\psi}\ti{\psi})\right] $ and make replacements as  $U\rightarrow \delta U$, $\gamma\rightarrow \delta \gamma$, $\gamma'\rightarrow \sqrt{\del}\gamma' $ in Eqs.\,(A1). Then, after writing $\ti{\psi} $, $\ti{\psi}^+$ in Cartesian form as
\begin{subequations}
	\begin{align}
	\ti{\psi}&= \frac{1}{\sqrt{2}}(\psi_1 + i \psi_2).\\
	\ti{\psi}^+&= \frac{1}{\sqrt{2}}(\psi_1 - i \psi_2).
	\end{align}
\end{subequations}
the total action may be rewritten as follows \cite{andersen}
\begin{subequations}
	\begin{align}
	S& =S_0 + S_{free}+S_{int}\label{eq:RS}\\
	S_{free}&= \frac{1}{2}\int_{0}^{\beta}d\tau\int d\vec{r}\left\lbrace i \epsilon_{ab} \psi_a \partial_\tau \psi_b +\psi_1 (-\hat{K}+ X_1)\psi_1 + \psi_2(-\hat{K}+ X_2)\psi_2 \right\rbrace.  \label{eq:RSfree} \\
	S_{int}&= S_{int}^{(1)} + S_{int}^{(2)} + S_{int}^{(3)} + S_{int}^{(4)}.\label{eq:RSint}\\
	S_{int}^{(1)}&= \int_{0}^{\beta}d\tau\int d\vec{r}\left\lbrace \delta \alpha_1 \psi_1 + \delta \alpha_2 \psi_2 -\sqrt{2}\gamma' \psi_2 \sqrt{\delta}\right\rbrace,  \label{eq:RSint(1)} \\
	S_{int}^{(2)}&=\frac{\delta}{2}\int_{0}^{\beta}d\tau\int d\vec{r} \left\lbrace \beta_1 \psi_1^2 + 2\beta_{12} \psi_1 \psi_2 + \beta_2 \psi_2^2\right\rbrace , \label{eq:RSint(2)} \\
	S_{int}^{(3)}&= \delta \int_{0}^{\beta}d\tau\int d\vec{r} \left\lbrace (\psi_1^2 +\psi_2^2)(\psi_1\gamma_1 + \psi_2\gamma_2)\right\rbrace, \label{eq:RSint(3)}\\
	S_{int}^{(4)}&=\frac{\delta U}{8}\int_{0}^{\beta}d\tau\int d\vec{r}\left\lbrace \psi_1^4 +2\psi_1^2 \psi_2^2 +\psi_2^4 \right\rbrace, \label{eq:RSint(4)}
	\end{align}
\end{subequations}
where
\begin{subequations}
	\begin{align}
	\beta_1& = -\mu -X_1 +\gamma +\frac{U\rho_0}{2}(\xi^2+ \bar{\xi}^2 + 4),\\
	\beta_2& = -\mu -X_2 -\gamma -\frac{U\rho_0}{2}(\xi^2+ \bar{\xi}^2 - 4),\\
	\beta_{12}& = \frac{iU\rho_0}{2}(\bar{\xi}^2 - \xi^2),\\
	\alpha_1&=\frac{2\gamma_1}{U} (-\mu +\rho_0 U +\gamma),\\
	\alpha_2&=\frac{2\gamma_2}{U} (-\mu +\rho_0 U -\gamma) -\sqrt{2}\gamma',\\
	\gamma_1& = \frac{U\sqrt{2\rho_0}}{4} (\bar{\xi} + \xi),\\
	\gamma_2&=  \frac{iU\sqrt{2\rho_0}}{4} (\bar{\xi} -\xi),
	\end{align}
\end{subequations}
and $X_1 =\Sigma_{n}+\Sigma_{an} -\mu$, $X_2 =\Sigma_{n}-\Sigma_{an} -\mu$ are the variational parameters. The free energy $\Omega$ can be evaluated as
\bea
\Omega&= -T \ln Z(j_1, j_2)|_{j_1=0, j_2=0},
\eea
where the grand partition function is
\bea
Z(j_1, j_2)= e^{-S_0}\int D \psi_1 D \psi_2 e^{-\frac{1}{2}
	\int dx \int dx' \psi_a(x)G_{ab}^{-1}(x,x')\psi_b(x')} e^{-S_{int}}
e^ { \int dx [j_1(x)\psi_1(x)+j_2(x)\psi_2(x)] }\nn
\label{eq:z}
\eea
in which we introduced  $x=(\tau,\vec{r})$ and $\int dx\equiv \int_{0}^{\beta}  d\tau\int d\vec{r}$.
For a uniform system, Green function is translationally invariant
\bea
G_{ab}(\vec{r}, \tau;  \vec{r}', \tau') = \frac{1}{\beta}\sum_{n,k} e ^{i\omega_n(\tau-\tau')} e^{i\vec{k}(\vec{r}-\vec{r}')} G_{ab}(\vec{k}, \omega_n)
\label{eq: Gab}
\eea
with
\begin{subequations}
	\begin{align}
	G_{11}(\vec{k}, \omega_n)&=\frac{\epsilon_k+X_2}{\omega_n^2+E_k^2},\\
	G_{22}(\vec{k}, \omega_n)&=\frac{\epsilon_k+X_1}{\omega_n^2+E_k^2},\\
	G_{12}(\vec{k}, \omega_n)&=\frac{\omega_{n}}{\omega_n^2+E_k^2},\\
	G_{21}(\vec{k}, \omega_n)& =-G_{12}(\vec{k}, \omega_n),\\
	E_k^2&=(\epsilon_k + X_1) (\epsilon_k + X_2)
	\end{align}
\end{subequations}
and $\epsilon_k$ is the bare dispersion \cite{23 our aniz}. In the path integral formalism the expectation value of an operator $\langle\hat{O}(\ti{\psi}^{+},\ti{\psi})\rangle$ is defined as
\bea
\langle\hat{O}\rangle = \frac{1}{Z_0}\int D \ti{\psi}^{+} D \ti{\psi} \hat{O}(\ti{\psi}^+,\ti{\psi}) e^{-S(\ti{\psi}^{+},\ti{\psi})},
\eea
where $Z_0=Z(j_1=0, j_2=0, S_{int}=0)$ is the noninteracting partition  function. Particularly, using the well-known formula \cite{Faddeev}:
\bea
\int D\psi_1 D \psi_2 exp{\left[ -\frac{1}{2} \sum_{a,b= 1, 2} \int\psi_a(x)G_{ab}^{-1}(x,y)\psi_b(y)dx dy + \int j_1(x)\psi_1(x)dx + \int j_2(x)\psi_2(x)dx\right]}  \nn
= (\sqrt{{\rm Det}\,G}) exp\left[\sum_{a,b= 1, 2}\int dx dy j_a(x) G_{ab}(x,y)j_b(y)\right] \nn
\label{eq: A10}
\eea
one may show that \cite{ouryeeint}
{
\begin{subequations}
	\begin{align}
	\langle \hat{O}(\psi_a(x)\psi_b(y))\rangle&=\hat{O} \left( \frac{\delta}{\delta j_a(x)}, \frac{\delta}{\delta j_b(y)}\right)  exp{\left[ \frac{1}{2} \int j_a(x) G_{ab}(x,y)j_b(y)dx dy\right] }, \\
	\langle \psi_a(x)\psi_b(x')\rangle&= G_{ab}(x,x'),\\
	\langle \psi_1(x)\psi_2(x) \rangle & = G_{12}(0) = \frac{1}{\beta}\sum_{n}G_{12}(\vec{k}, \omega_n) = \frac{1}{\beta} \sum_{n=-\infty}^{\infty}\frac{\omega_{n}}{\omega_{n}^2 + E_k^2}=0,\\
	\langle\psi_a^4(x)\rangle &= 3 G_{aa}^2(0),\\
	\langle\psi_1^2 (x) \psi_2^2 (x)\rangle& = G_{11}(0) G_{22}(0),\\
	G_{ab}(0)&\equiv\frac{1}{\beta} \sum_{k,n} G_{ab}(k, \omega_{n}),\\
	\langle \psi_{a_1}, \psi_{a_2}\ldots\psi_{a_n}\rangle&=0, \quad n=1,3,5 \ldots
	\end{align}
\end{subequations}
}
We now expand $exp(-S_{int})$ in Eqs.\,(A3) in powers of $\delta$
\bea
e^{-S_{int}}= 1-S_{int}^{(1)} - S_{int}^{(2)} - S_{int}^{(3)} - S_{int}^{(4)} +\frac{1}{2}[S_{DM}']^2 +O(\delta^{3/2})
\eea
where $S_{int}^{(i)}$ are given in Eqs.\, {(A.3)} and
{
\bea
S_{DM}'=-\gamma'\sqrt{2\delta} \int dx \psi_2(x).
\eea
}
Expressing the ``noninteracting" partition function as
\bea
Z_0(j)&=\int D \psi_1 D \psi_2 e^{-\frac{1}{2}\int dx dx'
	\psi_a(x)G_{ab}^{-1}(x,x')\psi_b(x')} e^{\int dx j_a(x)\psi_a(x)}\\
&=(\sqrt{DetG}) \exp
{\left[\frac{1}{2}\int dx dx' j_a(x) \bar{G}_{ab}(x,x')j_b(x') \right]}
\eea
where $\bar{G}_{ab}(x,y)=[G_{ab}(x,y)+G_{ba}(y,x)]/2$,  one may obtain
\bea
Z(j)&= e^{-S_0} \left[Z_0(j)-\langle S_{int}^{(1)}\rangle - \langle S_{int}^{(2)}\rangle - \langle S_{int}^{(3)} \rangle- \langle S_{int}^{(4)}\rangle +{\delta\gamma'^2 }
\int dx dx'\langle\psi_2(x) \psi_2(x')\rangle +O(\delta^{\frac{3}{2}}) \right] \nn
\label{eq:z(j)}
\eea
where $\langle\hat{O}\rangle=[\int  D \psi_1 D \psi_2 e^{-S_{free}}\hat{O}(\psi_1, \psi_2) ]/Z_0(j)|_{(j=0)} $ and $Z_0(j)|_{(j=0)}= 1/\sqrt{{\rm Det}\,G^{-1}}$. The expectation values in (\ref{eq:z(j)}) can be easily calculated by
Eqs.\,(\ref{eq: A10}) and {(A.11)}
{
\begin{subequations}
	\begin{align}
	\langle S_{int}^{(1)}\rangle& =0, \quad  \langle S_{int}^{(3)}\rangle =0, \\	\langle S_{int}^{(2)}\rangle &= \frac{1}{2}\int dx{(\beta_1 G_{11}(0) + \beta_2 G_{22}(0))}
\nn
	&=\frac{\beta}{2}({\beta_1 B +\beta_2 A})\\
	\langle S_{int}^{(4)}\rangle &= \frac{U\beta}{8}\left[ 3G_{11}^2(0)+3G_{22}^2(0)+2G_{11}(0)G_{22}(0)\right] \nn
	&= \frac{U\beta}{8}[3B^2 + 3A^2+2A B].\\
	\int dx dx' \langle\psi_2(x) \psi_2(x')\rangle&=\frac{1}{\beta} \sum_{n,k} \int d\vec{r} d\vec{r'}d\tau d\tau' e^{i\omega_{n}(\tau-\tau')} e^{i\vec{k}(\vec{r}-\vec{r}')}G_{22}(\vec{k}, \omega_n) \nn
	&=\beta G_{22}(\vec{k}, \omega_n)|_{(k=0, n=0)}=\frac{\beta(\epsilon_k+X_1)}{\omega_{n}^2+E_k^2}|_{(k=0, n=0)} \nn
	&= \frac{\beta X_1}{(\epsilon_k+X_1)(\epsilon_k+X_2)}|_{(k=0,n=0)} = \frac{\beta}{X_2}.
	\end{align}
\end{subequations}
}
where $A$, $B$, $\beta_{1,2}$ are given by Eqs.\,(16).
Thus, using the formula $\ln(1+x)\approx x$ we obtain
\bea
\Omega = -T \ln Z(j)|_{j=0} = -T \ln e^{-S_0} -T \ln Z_0 + T\langle S_{int}^{(2)}\rangle + T\langle S_{int}^{(4)}\rangle -\frac{\gamma'^2}{X_2}\nn
\label{eq: finalomega}
\eea
where we set $\delta=1$. Finally, using {Eqs.\,(A.17)} gives
\begin{subequations}
	\begin{align}
	\Omega&= \Omega_0 + \Omega_{free} + \Omega_2 + \Omega_4 + \Omega_{DMA}^{(2)}\\
	\Omega_0& = -\mu\rho_0 + \frac{U\rho_0}{2} + \frac{\gamma\rho_0}{2}(\bar{\xi}^2 +\xi^2)-i\gamma'(\bar{\xi}-\xi) \sqrt{\rho_0},\\
	\Omega_{free}& = \frac{1}{2} \sum_{k} (E_k -\epsilon_k) + T \sum_{k}ln(1-e^{-\beta E_k}),\\
	\Omega_2& =  \frac{1}{2} [\beta_1 B + \beta_2 A],\\
	\Omega_4& =\frac{U}{8}[3A^2 +3B^2 +2AB],\\
	\Omega_{DMA}^{(2)} &= -\frac{\gamma'^2}{X_2}.
	\end{align}
\end{subequations}
which is presented in section II.
{
Note that in the isotropic case one should accurately introduce
an additional Lagrange multiplier, $\mu_0$ to obtain following expression for $\Omega$
\be
\Omega_{ISO}=-\mu_0\rho_0+U\rho_0/2+\Omega_{free}^{ISO}+\Omega_{2}^{ISO}+\Omega_{4}^{ISO}
\ee
where expressions for  $\Omega_{free}^{ISO}$, $\Omega_{2}^{ISO}$ and $\Omega_{4}^{ISO}$ are
formally the same as in Eq.s (A.19) with $\gamma=\gamma'=0$.
}
 Moreover, it can be shown that the number conservation condition, $\langle \ti{\psi}\rangle=0$, i.e., $\langle {\psi}_1\rangle=0$ and $\langle {\psi}_2\rangle=0$ will lead exactly to Eqs.\,(\ref{eq:cos1}) and (\ref{eq:cos2}). Evidently, $\delta$-expansion method makes it easy to take into account the linear interaction $H_{DMA}=i\gamma' \int d\vec{r}(\psi -\psi^+)$ to higher orders in $\gamma'$.

\end{document}